\documentclass[preprint,review,12pt]{elsarticle}

\usepackage[usenames,dvipsnames]{xcolor}
\usepackage{tikz} \usetikzlibrary{calc, arrows.meta, intersections, patterns, positioning, shapes.misc, fadings, through,decorations.pathreplacing}
\definecolor{ColorOne}{named}{MidnightBlue}
\definecolor{ColorTwo}{named}{Dandelion}
\definecolor{ColorThree}{named}{Plum}
 \usepackage{epsfig}
\usepackage[utf8]{inputenc}
\include{Configurations/Color}
\include{Configurations/HyperRefStyle}
\include{Configurations/Coding}
\usepackage{graphicx}

\usepackage{amssymb}
\usepackage{amsmath}
\usepackage{caption}
\PassOptionsToPackage{hyphens}{url}
\usepackage[hidelinks]{hyperref}
\usepackage{graphicx} 
\usepackage{amsmath,amsfonts}
\usepackage{algorithmic}
\usepackage{algorithm}
\usepackage{array}
\usepackage[caption=false,font=normalsize,labelfont=sf,textfont=sf]{subfig}
\usepackage{textcomp}
\usepackage{stfloats}
\usepackage{verbatim}
\usepackage{graphicx}
\usepackage{amsmath,amssymb,amsfonts}
\usepackage{algorithm}
\usepackage{graphicx}
\usepackage{textcomp}
\usepackage{multirow}
\usepackage{makecell}
\usepackage{mathtools}  
\usepackage{mathrsfs}   
\usepackage{amsmath,amssymb,amsfonts}
\usepackage{algorithm}
\usepackage{graphicx}
\usepackage{textcomp}
\usepackage{multirow}
\usepackage{makecell}
\usepackage{placeins} 

\newcommand\ci{\mathcal{I}}		
\newcommand\cj{\mathcal{J}}		
\newcommand\cb{\mathcal{B}}		
\newcommand\cl{\mathcal{L}}		

\DeclareMathOperator*{\argmin}{argmin}

\journal{Energy and AI}

\begin{document}


\begin{frontmatter}



\title{Integrated Learning and Optimization for Congestion Management and Cost Minimization in Real-Time Electricity Market}
\author[CEMSE]{Imran Pervez}\ead{imran.pervez@kaust.edu.sa} 
\author[CEMSE]{Ricardo M. Lima}\ead{ricardo.lima@kaust.edu.sa} 
\author[CEMSE]{Omar Knio\corref{cor1}}\ead{omar.knio@kaust.edu.sa} 
\affiliation[CEMSE]{organization={Computer, Electrical, and Mathematical Sciences and Engineering Division,King Abdullah University of Science and Technology},
            city={Thuwal},
            postcode={23955}, 
            state={Makkah},
            country={Saudi Arabia}}
\cortext[cor1]{Corresponding author.}
\begin{abstract}
We propose a novel integrated learning and optimization (ILO) methodology to predict contextual-driven parameters occurring in the DC optimal power flow (DCOPF) optimization problem used in real-time market (RTM) application. We also formulated ILO for contextual-driven parameter training in economic dispatch (ED) optimization problem for RTM application. These are critical optimization problems that are solved multiple times per day. Here, they are parameterized over the load demand and the power transfer distribution factor matrix that need to be predicted as a function of the weather conditions and time of the day, i.e., the context. The proposed methodology equips a neural network prediction model with two new features. The first is the inclusion of the DCOPF problem into the training algorithm, whereby optimal decisions are calculated for predictions. The second is a novel regret function in the training algorithm for DCOPF parameter training, which replaces a traditional loss/regret function, to measure the cost of the optimal decisions for each prediction. The main goal is to train prediction models that minimize the post-optimization regret rather than the prediction accuracy. We derive two regret functions and their gradients and establish the connection between the regret functions of the ED and DCOPF problem, and electricity real-time markets. The computational results show the superior performance of the integrated learning and optimization compared to a sequential learning and optimization framework in terms of lower correction costs in the real-time electricity market and lower transmission line congestion.
\end{abstract}





\end{frontmatter}



\section{Introduction}\label{sec:intro}\noindent
The shift towards renewable energy sources comes with several challenges like high rate of change of frequency, netload (demand - renewable generation) below base load, the need for fast ramping generators, relative high capacity of non-dispatchable sources, forecasting/prediction challenges, and reduced revenue recovery (market price - production cost) of conventional generators leading to significantly high production prices during peak or mid-peak hours. In addition, the load demand exhibits increasing unpredictability, eg., electric vehicles (EVs) bring new consumption profiles, and increasing loads due to massive data centers and industry electrification. 

The above-mentioned challenges 
need to be addressed for an economic and reliable operation of power systems. At the heart of economic and reliable power systems operations lies economic dispatch (ED) and DC optimal power flow (DCOPF) optimization problems. 
These problems generate power-dispatching decisions using various predictions in their optimization setting, like renewable generation, load demand, and power transfer distribution factors (PTDF). 

One of the crucial application of power systems is the electricity market mechanism that decides production and consumption schedules and their pricing. The electricity market operations are largely dependent on ED/DCOPF decisions which are based on several predictions, e.g, load demand and renewable technologies output. 
Inaccuracy in these predictions propagates through the optimization problems, leading to incorrect dispatch decisions, which may significantly impact market operations and revenues. 
%

Electricity markets are of various types like day-ahead market (DAM), intraday market (IDM), or real-time market (RTM). The current work mainly deals with DAM and RTM where the former involves scheduling market decisions prior to real-time operation and depends on different predictions while the latter involves scheduling real-time market operations. 

Since the DAM or other future market transactions depend on weather, load, renewable and other predictions, actual production and consumption from renewables and loads respectively is likely to deviate from market-cleared schedules in real-time, resulting in a supply-demand imbalance, or line congestion. 
%

Independent system operators (ISOs) use RTM mechanisms to balance supply and demand in near to/or real-time to avoid user inconvenience, blackouts, or unplanned demand cuts. 
The IDM and BM mechanisms solve ED/DCOPF problems in real-time to make purchasing/selling decisions for electricity from different market traders willing to generate/consume extra to ramp up/down the total system power for total supply and demand balancing and line congestion improvement. 
Deviations from nominations, due to inaccuracies in predictions, come at higher prices 
thereby increasing the total system cost.

In general, the ED or DCOPF problems are addressed in three phases: i) training phase: where a prediction model is trained using historical data including context and parameters required within ED and DCOPF; ii) prediction phase: given a specific context, the prediction model is applied to predict parameters for the ED or DCOPF problems; and iii) optimization phase: the ED or DCOPF problems, parameterized over the parameters determined in the prediction phase, are solved.
This is known as sequential learning and optimization (SLO) \citep{b30}. 
An alternative approach follows the three phases of SLO but in the training phase is augmented with a) the optimization problem from the optimization phase; and b) a loss function to measure deviations based on post-optimization decisions. 
This approach is known as integrated learning and optimization (ILO) \citep{b30}. 
The application of ILO to ED and DCOPF and its performance analysis is the main subject of this work. 
Through the text, these concepts, details, and methodology are further elaborated and the superiority of the ILO over the SLO is demonstrated.
\subsection{Literature overview}\label{sec:overview}\noindent
The literature overview covers three main areas relevant for this work. 
The first centers on learning algorithms developed to predict relevant parameters within power systems applications. 
The second covers stochastic optimization approaches using model predictive control to address ED problems. 
Finally, we focus on the subject of this work–ILO developments and applications. 
These three areas provide relevant tools to address prediction of parameters, and solution of optimization problems that address critical decision-making problems that involve assuming values for parameters that are unknown and depend on a context.

Various supervised learning algorithms have been proposed in the literature for training predictions in the ED/DCOPF optimization setting. 
In~\cite{b1}, the authors used a Levenberg–Marquardt back-propagation (LM-BP) neural network (NN) to improve the load prediction accuracy. This is achieved by combining the gradient descent and Quasi-Newton method to ensure faster speed, accuracy, and stability. The authors in~\cite{b2} used auto-machine learning (auto-ML) for feature extraction and look-ahead window (prediction horizon) size selection for look-ahead (multi-time) load forecasting. They show that the overall forecasting/prediction accuracy can be improved by optimizing hyperparameter selection. The authors in~\cite{b3} proposed a transfer learning- and temporal fusion transformer (TFT) approach to train prediction models for users and buildings with small load consumption data sets. The trained model provides significant improvement in accuracy compared to existing approaches despite less data availability. 

The authors in~\cite{b4} used a deep neural network with an unsupervised learning algorithm for feature extraction from the data. In~\cite{b5}, a deep convolutional neural network (DCNN) was proposed which uses convolutional layers and extra dense layers to generate accurate load prediction accuracy. In~\cite{b6}, the authors proposed a hybrid algorithm that uses convolutional neural network (CNN) to establish the load trend learning capability and long short term memory (LSTM) to capture patterns from the time-series data. 
A Gaussian process (GP) based load prediction method was proposed in~\cite{b7}. The proposed GP regression model employs compositional kernels to deal with the high dimensional data by giving weightage to the most important input features. Training with the most important features addresses high data dimensionality and enhances the overall prediction accuracy. 

The authors in~\cite{b8} proposed a homogeneous ensemble-based method (random forest (RF)) for load prediction. 
Moreover, the authors also analyzed 
the importance of different features in the dataset. The overall RF algorithm outperformed other algorithms in terms of accuracy. The authors in~\cite{b9} used a gradient boosting (GB) algorithm which iteratively combines several weak models (less accurate) to obtain an additive model using numerical optimization that minimizes the loss/regret function. The proposed method was found to improve the load prediction accuracy. The authors in~\cite{b10} proposed a vector field-based support vector regression method that maps a high-dimensional feature space using a vector field to find the optimal feature space. 
The proposed algorithm improved the accuracy and robustness of the load prediction. 
Though the above-mentioned methods improved the prediction accuracy
, none of them trained the predictions to learn better decisions on parameters that minimizes a regret function.
The literature also includes stochastic optimization techniques where supervised learning methods are used to learn predictions in the stochastic setting of ED/DCOPF models. 

The authors in~\cite{b11} used stochastic model predictive control (MPC) for distributed non-linear multi-objective ED which uses data-driven scenario generation using dynamic programming by including uncertain realizations from energy price, availability of renewable resources, and demand. Another work in~\cite{b12} used stochastic MPC using data-driven scenario generation with dynamic programming for centralized ED at different time horizons including uncertainty realizations from energy price, availability of renewable resources, and demand behavior. 

In~\cite{b13}, the authors developed centralized stochastic MPC using Monte Carlo (MC) simulation and Roulette wheel mechanism (RWM) for a contingency-constrained demand response (DR) problem by including uncertainty realizations from the availability of renewable resources, and the demand behavior. The authors in~\cite{b14} developed a centralized stochastic MPC using probability distribution functions (PDFs) for multi-objective ED with several different kinds of energy sources in a microgrid, namely by including uncertainty realizations from the availability of renewable resources. In~\cite{b15}, the authors proposed a centralized stochastic MPC method using the Markov chain Monte Carlo (MCMC) method for a multi-objective DR problem by including uncertainty realizations from the availability of wind resources and consumer demand. 

The authors in~\cite{b16} proposed centralized stochastic economic MPC for non-linear ED with balance responsible parties (BRPs) using Monte Carlo for scenario generation by including uncertainty realizations from the availability of wind resources. The authors in~\cite{b17} proposed a centralized stochastic economic hybrid with reference tracking (ERT) MPC with multi-objective ED formulation using sampling-based scenario generation by including uncertainty realizations from energy price, renewables, and demand. In~\cite{b18}, the authors proposed a stochastic ERT MPC for multi-objective ED using a sampling-based scenario generation approach by including uncertainty realizations from energy price and demand.

The authors in~\cite{b19} proposed a centralized stochastic economic MPC for ED using Gaussian process (GP) regression for uncertainty propagation. In~\cite{b20}, the authors proposed a centralized stochastic economic MPC for non-linear multi-objective ED problem using scenario-based uncertainty propagation by including electricity prices, weather, and demand as uncertain variables. 

The above-mentioned techniques used different stochastic MPC programming methods for various types of ED objectives. The application of stochastic MPC methods in all the above works outperformed the deterministic MPC and improved the ED optimum decision. However, the stochastic MPC and stochastic-robust MPC methods require significant computational efforts which may hamper their real-time use. Moreover, the stochastic methods are SLO-based and do not incorporate means to correct decisions so as to train the predictions favouring better decisions.

An alternative approach to predict unknown parameters is the ILO methodology that focuses on training prediction models to minimize the cost of inaccurate decisions, rather than inaccurate predictions. 
%
The application of ILO to optimization problems, and specifically to energy related problems, is scarce in the literature.

In~\cite{b26}, the authors formulated a differentiable regret function for the problems of type shortest path whose regret function resembles a zero-one loss and is non-differentiable for ILO training. The authors set the convex upper-bound on the non-differentiable regret function to make it differentiable and named their training framework as smart predict then optimize plus (SPO+). The authors in~\cite{b27} proposed an interior point (IP) algorithm-based gradient calculation using logarithmic barrier relaxation for the non-differentiable ILO loss functions of the 0-1 knapsack problem, unit commitment, and shortest path problem. The above-mentioned works mainly used ILO to train unknown  parameters in the objective function.

The authors in~\cite{b28} proposed to train unknown  parameters in the constraints using an IP-based gradient calculation of loss function. The authors proposed generic gradient formulations for two categories of linear programs (LPs), namely packing LP and covering LP. The authors applied their gradient formulation to train unknown parameters in the max flow transportation problem, an alloy production problem, and a fractional knapsack problem. In~\cite{b28a}, the authors proposed an end-to-end wind power prediction method to optimize the energy system by estimating wind predictions to optimize decisions rather than wind prediction accuracy. In~\cite{b28b}, the authors used decision-focused learning for combinatorial optimization; however, their learning approach was based on decision rule optimization (DRO) which parametrizes both unknowns in the optimizaion model and the decision/policy. The decision/policy being an unknown and learned integratedly with another unknown in the optimization model may not represent the true decision/policy. The authors in~\cite{b28c} proposed an end-to-end approach using an energy based model to avoid calculating the gradient of the regret function at each epoch and used the approach for load prediction in power systems.

None of the works cite above formulated an ILO for DCOPF with multiple unknown parameters (load and PTDF predictions) in the constraints,  
considering their correlations, sensitivity analysis, their impact at different stages of market applications and ILO training formulations.

In this work, to the best of our knowledge, we develop novel ILO formulation to achieve an economic operation in real-time market operations and non-real time generator scheduling together. The ILO formulation is designed to capture the real-time market and generator scheduling procedures to be used as a feedback for training certain unknown parameters in the constraints of ED/DCOPF formulations. The ILO formulation for ED is developed to obtain mainly the real-time cost-effective market operations whereas a new ILO formulation for DCOPF is developed to obtain hour-ahead and real-time cost-effective economic market operations. The training and testing results of ILO are compared with the SLO based training of ED/DCOPF unknowns and the results demonstrate a significant performance of ILO over SLO. 
\section{Optimal dispatch and real time market operations}\label{sec:system}\noindent
The mathematical formulations for optimal dispatch as well as real-time market operations considered in this work are described in this section. 
\subsection{Economic Dispatch and DC Optimal Power Flow}\label{sec:ed-dcopf}\noindent
The ED optimization problem addresses the optimum resources allocation to serve consumer electric demand while meeting different system-wide constraints at minimum generation cost. 
The ED optimization problem is formulated as 
\begin{subequations}
\begin{align}
	\min_{p,\, s}\,\, 	& c^{\top} p + c^{\textrm{ext}\top} s, \\
	\textrm{s.t.} \,\, 	& 1^{\top} p + s = 1^{\top} d, \\
				& p \le p^{\max}, \\
				& -p \le - p^{\min}, \\
				& -s \le 0, 
\end{align}%
\label{eq:ed}%
\end{subequations}%
where $p \in \mathbb{R}^{|\ci|}$ denotes the power output of generators, $s \in \mathbb{R}$ is the additional power required from an external system, $c \in \mathbb{R}^{|\ci|}$ represents the generation costs, $c^{\textrm{ext}} \in \mathbb{R}$ is the cost of the external power, $d \in \mathbb{R}^{|\cb|}$ represents the load at each bus, $1$ is a vector with ones with the proper dimension, and $p^{\min}$, $p^{\max} \in \mathbb{R}^{|\ci|}$ represent the minimum and maximum generator power output limits, respectively. 

A more comprehensive formulation including lines capacities and power flows limits is known as DCOPF:
\begin{subequations}
\begin{align}
	\min_{p,\, s}\,\, 	& c^{\top} p + c^{\textrm{ext}\top} s, \\
	\textrm{s.t.} \,\, 	& 1^{\top} p + s =  1^{\top} d, \\
				& T(Mp + Ns - d ) \le p^{\textrm{line,max}}, \\
				& - T(Mp + Ns - d ) \le -p^{\textrm{line,min}}, \\
				& p \le p^{\max}, \\
				& -p \le - p^{\min},\\
				& -s \le 0, 
\end{align}%
\label{eq:dcopf}%
\end{subequations}%
where $T \in \mathbb{R}^{|\cl|\times|\cb|}$ is the PTDF matrix, $M \in \{0,1\}^{|\cb|\times|\ci|}$, $N \in \{0,1\}^{|\cb|\times|\cb|}$ are incidence matrices mapping generators to buses and interconnections with neighbor systems to buses, respectively, which are described and in several works~\cite{b28h}-\cite{b28i}, and $p^{\textrm{line,min}}$, $p^{\textrm{line,max}}  \in \mathbb{R}^{|\cl|}$ are vectors with the minimum and maximum lines capacities, respectively. 

Problem \eqref{eq:ed} is solved for each hour of the day, where each hour is characterized by a specific context, e.g., the weather conditions, the time of the day, the day of the year, or other relevant features. 
The context influences the load demand, and therefore, the load demand needs to be predicted as a function of the context for each hour. 
In addition, we assume that in Problem  \eqref{eq:dcopf}, the PTDF matrix depends on the context, and it needs to be predicted before solving the DCOPF. The idea of predicting PTDF matrix comes from the flexible AC transmission systems (FACTS) devices that provide the flexibility to change line impedances of a transmission network~\cite{b28d}-\cite{b28g}. The PTDF matrix coefficients directly depend on the transmission line impedances and system topology and thus varies with line impedance variations through FACTS devices. 
We denote the parameters that depend on the context as unknown parameters, while  $c^{\top}$, $c^{\textrm{ext}}$, $p^{\min}$, $p^{\max}$, $p^{\textrm{line,min}}$ and $p^{\textrm{line,max}}$ do not depend on the context, and thus, they are denoted as known parameters.
\subsection{Real time market operations}\label{sec:rtm}\noindent
The current work studies the grid side operations where an ISO balances supply-demand deviations from market clearing schedules in RTM and minimize line-congestion by ramping-up and -down generators. 
Prior to RTM, the DAM is cleared providing day-ahead set-points that along with other predictions provided by the ISO are applied in ED/DCOPF formulations to generate power dispatching decisions. 
The prediction-driven ED/DCOPF decision deviates from the true decision due to deviations in prediction. 
The deviations in load prediction mainly result in supply-demand imbalance while the deviations in PTDF prediction lead to line congestion. 
The ISO corrects deviations in prediction-driven decisions by interacting with real-time market participants willing to ramp-up/-down their generators for supply-demand balancing. 
Moreover, the ISO can also control line impedances to create a correlation of impedances between lines which depends on load and system topology to minimize line congestion. 
The ISO solves the ED/DCOPF problems to supply consumer loads by scheduling different generators using load prediction while also controlling line impedance to generate PTDF predictions which minimize line congestion during generator scheduling. 
After the true load is known, the ISO corrects supply-demand imbalances by ramping-up or -down different generators in the market at prices different than the day-ahead market clearing price (MCP). 
Moreover, the effect of line congestion on real-time market costs is also minimized due to impedance control during prior to real-time scheduling. 
The ISO also trades electricity with different regions if the generators within the same region cannot ramp-up/-down corresponding to prediction inaccuracies. 

Due to price differences from MCP, the demand participants are subjected to extra payments (indirect penalty due to generator ramping) for both over and underestimations in the load. Moreover, if line impedances are not controlled/predicted well the corresponding PTDF predictions will lead to higher operational costs during hour-ahead scheduling and also higher ramping costs during real-time correction. The current study thus, instead of minimizing load and PTDF prediction errors, minimizes extra payments on demand participants as a result of generator ramping and line congestion due to inaccurate load and PTDF predictions using ILO. The ILO as explained in the upcoming sections integrates learning and optimization to capture ISO real-time correction procedures to be used as a feedback to learn the load and PTDF predictions to minimize extra payments on demand participants and line congestion. The functioning of the system under study is illustrated in Fig.~\ref{fig:ILO_ED_mark}.
\begin{figure}
    \centering
    \includegraphics[width=1.02\linewidth]{Figures/market_mech_opf___.jpg}
        \caption{Two markets namely the day ahead market (DAM) and the real-time market (RTM). The market operator (MO) interacts with market participants in the DAM to schedule load and generators for next 24 hours ahead of that day (shown in Fig., say at 8 Sept., 10 am the MO schedules the loads for the next 24 hours starting from 9 Sept., 2 pm). The predicted PTDF ($T_{\theta}$) matrix is obtained by line impedance predictions such that each line impedance is correlated to other lines to minimize line congestion. The true load ($d$) and true PTDF matrix ($T$) are realized in real-time and do not match the predicted load and PTDF. The ISO interacts with the market participants to correct load imbalance in real-time, while prior to real-time the ISO control line impedance to minimize line congestion. Inspired by the above-mentioned process, we use ILO training pipeline to train load and PTDF matrix predictions to minimize ramping-up($\uparrow$)/ramping-down($\downarrow$) costs, penalize $\uparrow \text{more than} \downarrow$, and minimize congestion}
    \label{fig:ILO_ED_mark}
\end{figure}
\FloatBarrier
\section{Integrated learning and optimization methodology}\label{sec:ilomethod}\noindent
A common approach to solve  Problems \eqref{eq:ed} and \eqref{eq:dcopf} with unknown parameters is based on three phases: an initial phase to train prediction models, a second phase to predict the unknown parameters, $d$ and $T$, based on the models trained in the first phase, and a third phase where the optimization problem is solved using those predictions. 
This approach is denoted as SLO because the three phases are solved in sequence without any feedback loop from the post-optimization outcomes to the prediction phase. 

Consider a linear program with the general formulation 
\begin{subequations}
\begin{align}
	z^*(y):=  & \min_{x}\,\, c^{\top} x, \\
	\textrm{s.t.} \,\, 	& g(x,y) \le 0,
\end{align}%
\label{eq:lp}%
\end{subequations}%
where $x \in \mathbb{R}^{n}$ is a vector of decision variables, $c \in \mathbb{R}^{n}$ is a vector of known parameters, $g: \mathbb{R}^{n}\rightarrow \mathbb{R}^{m}$ represents a set of $m$ constraints, $y \in \mathbb{R}^{m}$ is a vector of unknown parameters that depend on the context, and $z \in \mathbb{R}$ is parameterized over $y$. 
Problem \eqref{eq:lp} captures the structure of Problems \eqref{eq:ed} and \eqref{eq:dcopf} in terms of variables, constraints, and parameters.

In general, the prediction models in the initial phase implement a method that minimizes the prediction errors in respect to the true (or realized) values. 
For example, consider the prediction of the parameter $y$ that depends on a context represented by $A$ using a prediction model with the following components: \citep{b26}: 
\begin{enumerate}
	\item Historical training data is available represented by \{($A_{1}$, $y_{1}$), ..., ($A_{I}$, $y_{I}$)\}, mapping contextual information, $A_{i}\in\mathbb{R}^{m\times n}$, with the corresponding realizations of $y$.
	\item A hypothesis class $\mathcal{H}$ of prediction models $f: \mathcal{X}\rightarrow \mathbb{R}^{d}$, where the predicted unknown parameter, $\hat{y}$, is a function of the context: $\hat{y}:=f(A)$.
	\item A loss function $l:\mathbb{R}^{d}\times\mathbb{R}^{d} \rightarrow \mathbb{R}_{+}$ that quantifies the error of the prediction $\hat{y}$ in respect to the true value of $y$. For example, this loss function can be a least square loss function.
	\item Given the training data and the loss function, by the Empirical Risk Minimization (ERM) principle a prediction model $f^{*} \in \mathcal{H}$ is determined by 
	\begin{equation}
		f^{*}:= \argmin_{f\in\mathcal{F}} \frac{1}{I} \sum\limits_{i=1}^{I}{l[f(A_{i}),y_{i}]}.
		\label{eq:erm}
	\end{equation}
\end{enumerate}

In the sequential learning and optimization approach, after determining $f^{*}$ in the first phase, for a given context $\hat{A}$ the prediction $\hat{y}:=f^{*}(\hat{A})$ is calculated in the second phase, and finally the optimization problem \eqref{eq:lp} is solved using $\hat{y}$. 
%
%
%
%

\begin{figure}[!b]
        \begin{center}
    			\includegraphics[width=1\textwidth]{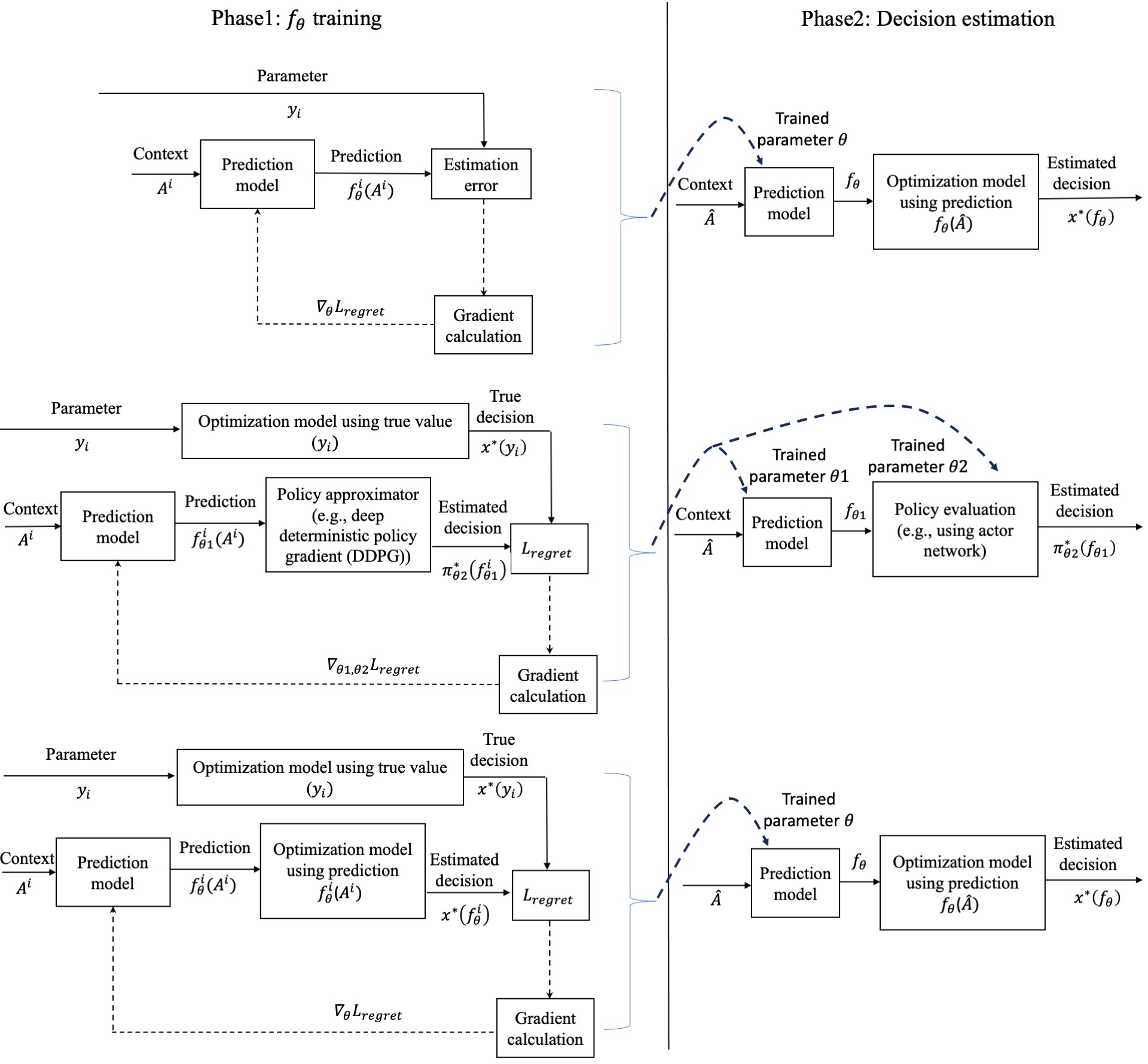}
		\end{center}
		\caption{Comparison between different training pipelines, (a) Sequential learning and optimization (SLO), (b) Integrated learning and optimization (ILO), and (c) Decision rule optimization (DRO)}
		\label{fig:pipelines}
\end{figure}

The SLO pipeline is summarized in Figure~\ref{fig:pipelines}(a) where for a given $i^{th}$  context $A^{i}$ the prediction output $f_{\theta}^{i}(A^{i})$ is calculated to minimize estimation error with respect to true value ($y$) using ERM principle to train model parameters $\theta$ by backpropagating gradient of the loss function. The trained $f_{\theta}^{i}(A^{i})$ is used to solve the optimization problem in the next phase making the pipeline sequential (predict and optimize sequentially) which is further quality checked using \eqref{eq:regret}.

In the SLO approach, in a post-optimization phase and after collecting the following information:
\begin{enumerate}
	\item the realization of $y$;
	\item the decisions $x^{*}(\hat{y})$ obtained from the optimization problem \eqref{eq:lp};
	\item the decisions $x^{*}_{corr}(\hat{y})$ implemented in reality; and 
	\item the optimum decisions $x^{*}(y)$ calculated after the realization of $y$;
\end{enumerate}
the prediction $\hat{y}$ can be assessed using a loss function (regret function) defined as \citep{b28,b29}:
\begin{equation}
	L(x^{*}(\hat{y}), y) := \underbrace{c^{\top}x^{*}_{corr}(\hat{y}) - z^{*}(y)}_{\text{Regret term}} + \underbrace{\phi \circ c^{\top}\left[x^{*}(\hat{y}) - x^{*}_{corr}(\hat{y})\right]}_{\text{Penalty term}},\\  
	\label{eq:regret}
\end{equation}
where $z^{*}(y):=c^{\top}x^{*}(y)$ is the minimum cost that could be achieved for the true $y$, and $x^{*}(\hat{y})$ is the solution obtained from problem \eqref{eq:lp}. 
The decisions $x^{*}_{corr}(\hat{y})$ implemented result from a correction applied to $x^{*}(\hat{y})$ to adapt to a true $y$ that is different from $\hat{y}$. 
Therefore, the first term captures the cost increase of implementing $x^{*}_{corr}(\hat{y})$ rather than $x^{*}(y)$, while the second term accounts for a penalty incurred to correct $x^{*}(\hat{y})$ to $x^{*}_{corr}(\hat{y})$; with a penalty denoted by $\phi$, multiplied element-wise by cost vector $c$.

In the SLO, this post-optimization information, i.e., the regret function value is not exploited to driving better predictions. 
However, it is a main component in ILO. Below, we distill the ILO framework providing notation and formulations based on \citep{b26,b27,b28}. 
\subsection{Integrated learning and optimization}\label{sec:ilo}\noindent
{\color{black} The proposed ILO methodology (illustrated in Figure \ref{fig:pipelines}(b)) extends the prediction method described in the previous section in three aspects: 1) the historical training data; 2) the integration of the optimization problem \eqref{eq:lp} in the prediction method; and 3) a novel regret function that replaces the loss function. 

The historical training data \{($A_{1}$,\, $y_{1}$), ..., ($A_{I}$,\, $y_{I}$)\} is augmented with $(x^{*}(y_i),\, z^{*}(y_{i}))$, where $x^{*}(y_i)$ represents the optimal decisions and $z^{*}(y_{i})$ the optimal solution obtained from \eqref{eq:lp} for ($A_{i}$, $y_{i}$). 
By integrating an optimization step into the prediction method, the optimal solution $(x^{*}(f_{\theta}^{i}),\, z^{*}(f_{\theta}^{i}))$ can be obtained from problem \eqref{eq:lp} for the prediction $f_{\theta}^{i}(A^{i})$. 
Based on this additional data\footnote{The historical pair $(x^{*}(y_i),z^{*}(y_{i}))$ and the optimal solution $(x^{*}(f_{\theta}^{i}),\, z^{*}(f_{\theta}^{i}))$.}, a new regret function based on \eqref{eq:regret} is implemented to replace the loss function $l$. 
This regret function quantifies the error of the optimal decisions $x^{*}(f_{\theta}^{i})$ in respect to $x^{*}(y_i)$, where $f_{\theta}$ is a prediction model parameterized over $\theta$. 
Note that in the sequential learning and optimization, the loss function quantifies the error of $f_{\theta}^{i}$ in respect to $y_{i}$.
To update $\theta$, the gradient of the regret function is evaluated and passed to the prediction algorithm.

The work in \citep{b26} formalized the ILO framework for cases with optimization problems with unknown parameters occurring only in the objective function. 
That is a particular case where the feasible region of the optimization problem does not depend on the unknown parameter, and therefore, the decisions obtained with the prediction are feasible for any realization of $y$. 
However, if the unknown parameters occur in the constraints, then the decisions obtained with the prediction may not be feasible for all realizations of $y$, and the loss function should account for the cost of the correction required to bring the decisions to the feasible region defined by the true $y$. 
The development of these corrections has been reported in \citep{b27,b28} for specific applications. Accordingly, we gave an example of a generic regret function in \eqref{eq:regret} which minimizes regret and corrects for infeasibility in the prediction-driven solution $x^{*}(\hat{y})$ with respect to the true solution ($x^{*}(y)$).  

After determining $f^{*}$, for a given context $\hat{A}$, the prediction $\hat{y}:=f^{*}(\hat{A})$ is used in the solution of the optimization problem \eqref{eq:lp}. 
As in the SLO framework, in a post-optimization step after the realization of $y$, the quality of the prediction can be quantified using the same regret function used within the prediction model. 
}

In this work, we propose two novel loss functions that include correction terms for the ED and DCOPF problems and establish a connection between the loss function, correction and the RTM described in Section \ref{sec:rtm}

\noindent
\underline{\textbf{Remark:}} The current work focus on the ILO and SLO pipelines; however, another pipeline namely decision rule optimization optimization (DRO) exists (see Fig.~\ref{fig:pipelines} (c)). The DRO pipeline trains the prediction model instances ($f_{\theta1}^{i}(A^{i})$) using policy approximation algorithms. The policy ($\pi_{\theta2}^{*}(f_{\theta1}^{i})$) contain parameters for prediction and policy models. The corresponding regret function ($L$) is minimized to update parameters $\theta1$ and $\theta2$ corresponding to prediction model and policy model respectively. 

Based on the above comparisons, the SLO pipeline learns the load prediction model to minimize error concerning the ground truth load. The load prediction model trained accurately used in optimization problem may not yield decisions favouring better power system operations despite good prediction accuracy.  

The closest in decision based learning concept to ILO is the DRO pipeline (eg., deep reinforcement learning (DRL)) which, similar to ILO, learns the prediction model to optimize decisions. However, DRO maps the context to an approximated/parameterized decision. The DRO thus involves two approximations, prediction of optimization problem unknowns and prediction of optimization problem decisions. The parametrized DRO decisions learned integratedly with other problem unknowns may not represent the true problem decisions.
\subsection{ILO applied to ED/DCOPF and RTM}\label{sec:iloeddcopf}\noindent
The concept of ILO can be extended to train unknown parameters in the ED and DCOPF optimization models using feedback information from real power system operations. The feedback information refers to some output of the power system in response to the prediction-driven decision input to the power system. The ED optimization model has the load demand (say $f_{true}$) (demand participant load) as an unknown parameter in the power-balancing equality constraints. The DCOPF has $f_{true}$ and PTDF matrix (say $PTDF$) as unknown parameters in its equality constraints. The unknown parameters as explained previously are the parameters in the optimization problem setting that depend on the context and need to be estimated/predicted using contextual information before solving the optimization problem. 
\subsubsection{Post-hoc (a posteriori) Analysis}\noindent
The load and PTDF estimation using ILO to minimize RTM operation costs and line congestion is based on the concept of post-hoc analysis. The post-hoc analysis involves correcting decisions corresponding to predictions using real-time true decisions. The post-hoc analysis resembles the real-world decision correction, where a decision taken prior to real-time using prediction differs from the decision made in real-time using true value. The difference between the two decisions is corrected in real-time, which, depending on the application may or may not incur extra correction penalty (extra cost). For model training using ILO, the correction between decisions with or without a penalty constitutes a regret function whose gradient is used to train the model parameters. 

For load training using ILO, the ED problem is initially solved using the load prediction model. Once the true load is known in real-time (historical load data during training), the ILO minimizes an objective-specific regret function ($L$) (minimizing RTM operation costs in the current study). The regret function is minimized by updating load prediction model parameters using gradient descent (GD) which eventually minimizes RTM operation costs as detailed in the upcoming sub-section. 

For load and PTDF training, the DCOPF problem is solved using load and PTDF predictions and corrected in real-time. The correction for DCOPF training contains two unknown parameters and thus require understanding of the impact of one parameter over the other to design the regret function as explained in the following subsection. The minimization of regret function to train load and PTDF matrix eventually minimizes line congestion and RTM operation costs. 
\subsubsection{Regret Function Design for ED and DCOPF Applied to RTM}\label{sec:regreted}\noindent
The $L$ for ED load training as explained previously is based on the concept of minimizing RTM operation costs. In the current study, the RTM operation costs are additional costs (higher than market costs) incurred as a result of generator ramping operations to balance supply-demand in real-time due to inaccurate load predictions. 
The ramping-up price (bidding price (BP)) 
and ramping-down price (offer price (OP)) 
must hold the following relationships with MCP: $BP>MCP$ and $OP<MCP$.
These pricing relationships indicate the extra price to be paid by the demand-side market participant for incorrect load estimations to the regulation market participants willing to ramp-up or -down their generation for supply-demand balancing. The regulation market players pay less when buying energy from ISO while charging more when selling energy to the ISO thereby imposing extra price/penalty to the demand participants. 
\begin{figure*}
    \centering
    \includegraphics[width=1.05\linewidth]{Figures/L_regret_.jpg}
    \caption{$L$ structure for ED/DCOPF unknown parameter training in equality constraints. The notation used for predictions and true parameters is same as used while explaining generic regret function formulations.}
    \label{fig:L_regret_struc}
\end{figure*}
\\
We propose a novel regret function based on \eqref{eq:regret} and on the following assumptions:
\begin{enumerate}
	\item $x^{*}(\hat{y})$ is corrected to optimality, whereby $x^{*}_{corr}(\hat{y}) = x^{*}(y)$, and therefore, $c^{\top}x^{*}_{corr}(\hat{y}) - z^{*}(y)=0$.
	\item  $x^{*}_{corr}(\hat{y}) = x^{*}(\hat{y}) + r + s$, where $r,\, s \in \mathbb{R}^{n}$ are correction factors needed to bring the system from $x^{*}(\hat{y})$ to $x^{*}_{corr}(\hat{y})$.
\end{enumerate}
The first assumption is motivated by the operation of a power system where the power generated is equal to the power consumed.  While in the context of the ED and RTM, $r := r^{+}-r^{-}$ where $r^{+},\, r^{-} \in \mathbb{R}^{n}_{+}$ represent the ramp-up and ramp-down power output of generators, , respectively, and $s := s^{+}-s^{-}$ where $s^{+},\, s^{-} \in \mathbb{R}^{n}_{+}$ are the additional power obtained and supplied to neighbor power systems, respectively.
With these assumptions and inspired on the extra price/penalty concept, the general structure of $L$ for ILO applied to ED and  RTM is defined as:
\begin{equation}
\label{eq7}
\begin{split}
L & = c^{\textrm{BP}\top}r^{-} + c^{\textrm{BP,ext}\top}s^{-} 
              	+ \left(c^{\textrm{MCP}}\oslash c^{\textrm{OP}}\right) \circ c^{\textrm{MCPT}} r^{+}
               	+ \left(c^{\textrm{MCP,ext}}\oslash c^{\textrm{OP,ext}}\right)\circ c^{\textrm{MCP,extT}}s^{+} \\
& \text{with the following constraints}\\
& r^{+}\geq p^{*}(f_{\theta})-p^{*}(f_{true}),\\
& r^{-}\geq p^{*}(f_{true})-p^{*}(f_{\theta}),\\
& s^{+}\geq s^{*}(f_{\theta})-s^{*}(f_{true}),\\
& s^{-}\geq s^{*}(f_{true})-s^{*}(f_{\theta}),\\
& r^{+}, \, r^{-} \geq 0,\\
& s^{+},\, s^{-}\geq 0,
\end{split}
\end{equation}%
where $c^{\textrm{BP}}$, $c^{\textrm{OP}}$, $c^{\textrm{MCP}}$ and $c^{\textrm{BP,ext}}$, $c^{\textrm{OP,ext}}$, $c^{\textrm{MCP,ext}}$ denotes prices to calculate the cost of power ramping-up at BP, cost of power ramping-down at OP, cost of power production at MCP within the same region and power ramping-up at BP, cost of power ramping-down at OP, cost of power production at MCP within the different region respectively. The symbols $\oslash$ and $\circ$ represent the hadamard element-wise division and multiplication of vectors respectively. 

The terms $c^{\textrm{BP}\top}r^{-}$~\text{and} $c^{\textrm{BP,ext}\top}s^{-}$
represent penalty on the demand participants when regulation market players selling power to ISO at higher than MCP. The terms $\left(c^{\textrm{MCP}} \oslash c^{\textrm{OP}}\right)\circ c^{\textrm{MCPT}} r^{+}$~\text{and} $\left(c^{\textrm{MCP,ext}}\oslash c^{\textrm{OP,ext}}\right)\circ c^{\textrm{MCP,extT}}s^{+}$ \ 
represent penalty to demand participants when regulation market participants purchasing power from ISO at less than MCP. The demand participants here refer to the market participants which provide load prediction prior to real-time. 

In both cases, demand participants will be penalized which in other words also means paying higher than MCP by a certain amount. The above relation thus in terms of cost due to MCP can be written as 
\begin{equation}
\begin{split}
L & :=  \phi^{+}c^{\textrm{MCP}\top}r^{+} 
	+ \phi^{+\textrm{ext}}c^{\textrm{MCP,ext}\top}s^{+} 
	+ \phi^{-}c^{\textrm{MCP}\top}r^{-}
	+ \phi^{-\textrm{ext}\top}c^{\textrm{MCP,ext}\top}s^{-} \\
& \text{with the following constraints}\\
& r^{+}\geq p^{*}(f_{\theta})-p^{*}(f_{true}),\\
& r^{-}\geq p^{*}(f_{true})-p^{*}(f_{\theta}),\\
& s^{+}\geq s^{*}(f_{\theta})-s^{*}(f_{true}), \\
& s^{-}\geq s^{*}(f_{true})-s^{*}(f_{\theta}),\\
& r^{+}, \, r^{-} \geq 0,\\
& s^{+},\, s^{-}\geq 0,
\end{split}
\label{{eq:regret}ED}
\end{equation}
where $\phi^{+}$ and $\phi^{+\textrm{ext}}$ represent penalty factors with respect to MCP for ramping-down (corresponding to OP), while the terms $\phi^{-}$ and $\phi^{-\textrm{ext}}$ represent the penalty factors with respect to MCP for ramping-up (corresponding to BP). 
\begin{figure*}
    \centering
    \includegraphics[width=0.9\linewidth]{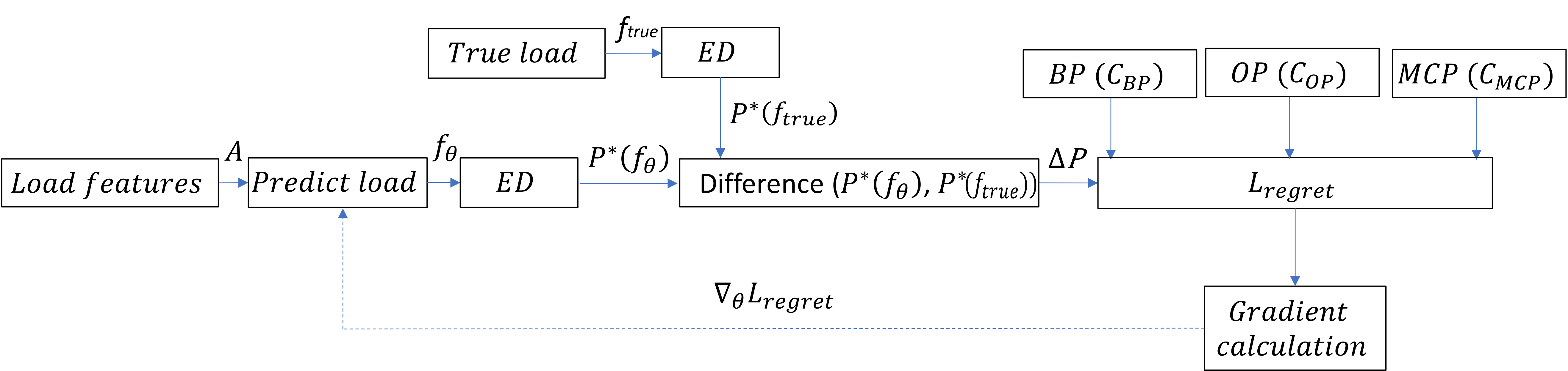}
    \caption{Load training using ILO. The predicted load ($f_{\theta}$) and the true load ($f_{true}$) are used to solve the ED problem to generate corresponding power set-points $p^{*}(f_{\theta})$ and $p^{*}(f_{true})$ respectively. The difference between the estimated and true power set-points along with the bidding price (BP), offer price (OP), and market clearing price (MCP)) are used to calculate the $L$ using~(\ref{{eq:regret}ED}). The gradient to~(\ref{{eq:regret}ED}) is calculated using procedures described in the next sub-section and backward passed to update $f_{\theta}$ parameters.}
    \label{fig:ILO_ED}
\end{figure*}
\begin{figure*}
    \centering
    \includegraphics[width=0.7\linewidth]{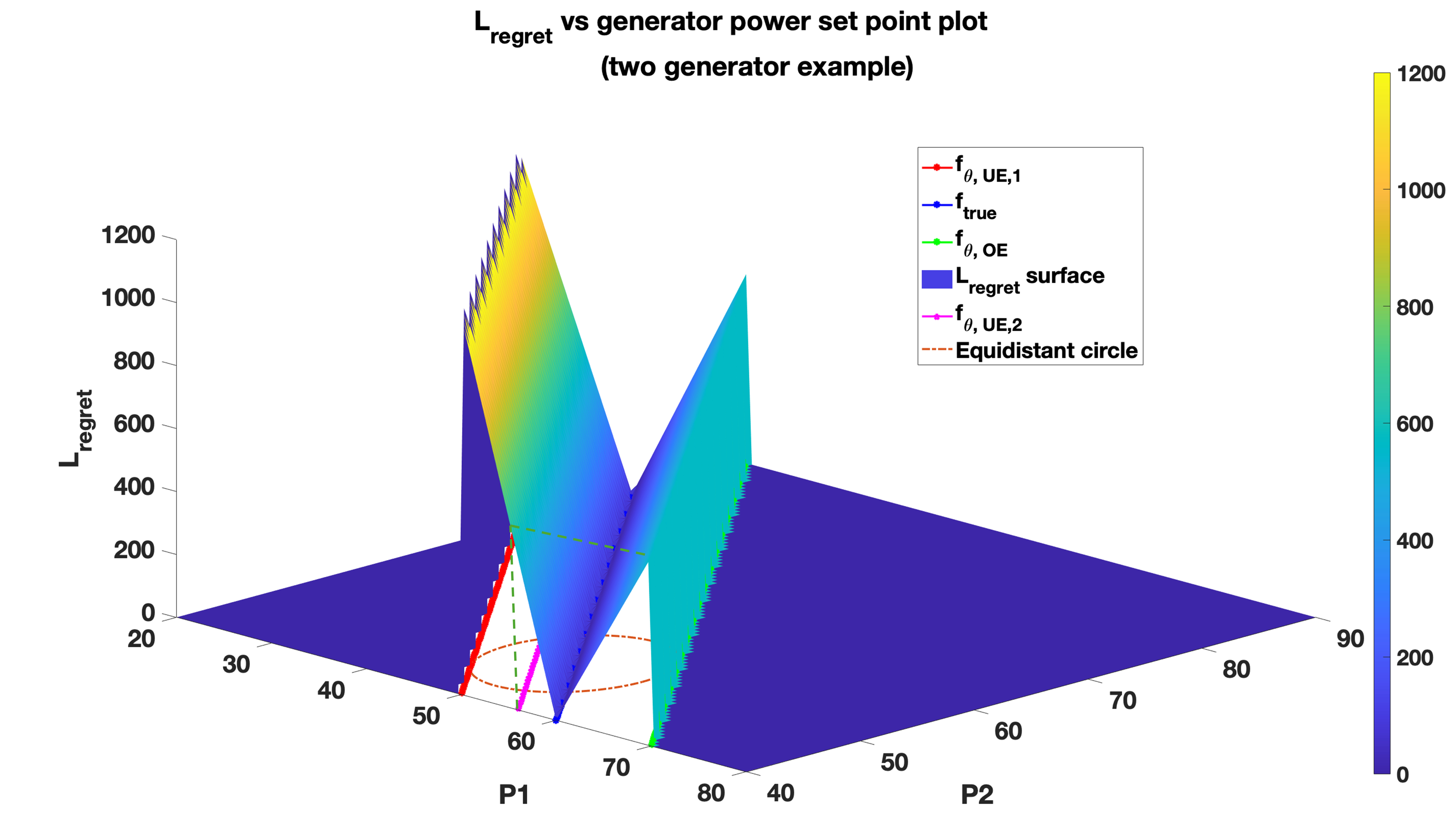}
    \caption{$L$ versus generator power set point ($\{P1, P2\}$) plot (two generator example). The plot represents $L$ surface corresponding to a load region in the x-y plane (P1-P2 space). The x-y plane represent feasible regions for different loads (overestimated, underestimated, and true loads). The overestimated load ($f_{\theta, OE}$) and underestimated load ($f_{\theta, UE, 1}$) load are equidistant from the true load ($f_{true}$) represented using equidistant circle. However, as illustrated the $L$ corresponding to $f_{\theta, UE, 1}$ is higher than the $L_{rgret}$ corresponding to $f_{\theta, OE}$. This is due to $L$ formulation to capture real-time market procedures of penalizing load underestimations more than overestimations. Another load underestimation within the region of interest ($f_{\theta, UE, 2}$) exhibits same $L$ to that of $f_{\theta, OE}$; however, much closer to $f_{true}$ compared to $f_{\theta, OE}$. The designed $L$ thus train the load to resemble either more overestimations or underestimations very close to the true load.}
    \label{fig:case1_feas_1}
\end{figure*}

Generally, the extra price for ramping-up is higher than the amount paid for ramping-down power generation. The $L$ thus, in addition to minimizing the ramping costs for incorrect load estimations is designed to penalize more load underestimations (require ramping-up) compared to load overestimations (require ramping-down). The $L$ concept for ED is illustrated for a two generator example in Fig.~\ref{fig:case1_feas_1}. 
\subsubsection{Gradient for the ED's regret function}\label{sec:edgradient}\noindent
To train the load, the $L$ in~(\ref{{eq:regret}ED}) will be minimized using the ERM principle~(\ref{eq:erm}) 
and the gradient descent (GD) algorithm, which requires calculating the gradient of $L$ in~(\ref{{eq:regret}ED}) with respect to unknown parameters $\theta$. To calculate the gradient, (\ref{{eq:regret}ED}) is rewritten as
\begin{equation}
\begin{split}
L  := & \ \phi^{+}c^{\textrm{MCP}\top}r^{+}\max\left[p^{*}(f_{\theta})-p^{*}(f_{true}), 0\right]\\
	+ & \ \phi^{+\textrm{ext}}c^{\textrm{MCP,ext}\top}s^{+}\max\left[s^{*}(f_{\theta})-s^{*}(f_{true}), 0\right] \\
	+ & \ \phi^{-}c^{\textrm{MCP}\top}r^{-}\max\left[p^{*}(f_{true})-p^{*}_{i}(f_{\theta}), 0\right]  \\
	+ & \ \phi^{-\textrm{ext}}c^{\textrm{MCP,ext}\top}s^{-}\max\left[s^{*}(f_{true})-s^{*}(f_{\theta}), 0\right].
\end{split}
\label{{eq:regret}EDmax}
\end{equation}
The differential to~(\ref{{eq:regret}EDmax}) using the total law of derivatives is calculated as
\begin{equation}
\begin{split}
\nabla_{\theta}L(f_{\theta},f_{true}) & := \frac{\partial L(f_{\theta},f_{true})}{\partial \theta}\\
	&= \frac{\partial L(f_{\theta},f_{true})}{\partial p^{*}(f_{\theta})}\frac{\partial p^{*}(f_{\theta})}{\partial f_{\theta}}\frac{\partial f_{\theta}}{\partial \theta} + \frac{\partial L(f_{\theta},f_{true})}{\partial s^{*}(f_{\theta})}\frac{\partial s^{*}(f_{\theta})}{\partial f_{\theta}}\frac{\partial f_{\theta}}{\partial \theta}\\
	&= 
\left\{\phi^{+}\frac{\partial c^{\textrm{MCP}\top}r^{+}}{\partial p^{*}(f_{\theta})}\frac{\partial p^{*}(f_{\theta})}{\partial f_{\theta}}\frac{\max[p^{*}(f_{\theta})-p^{*}(f_{true}),0]}{p^{*}(f_{\theta})-p^{*}(f_{true})} \right. \\& \left. 
+ \phi^{+\textrm{ext}}\frac{\partial c^{\textrm{MCP,ext}\top}s^{+}}{\partial s^{*}(f_{\theta})}\frac{\partial s^{*}(f_{\theta})}{\partial f_{\theta}}\frac{\max[s^{*}(f_{\theta})-s^{*}(f_{true}),0]}{s^{*}(f_{\theta})-s^{*}(f_{true})}\right\}\frac{\partial f_{\theta}}{\partial \theta}\\
	& 
+ \left\{\phi^{-}\frac{\partial c^{\textrm{MCP}\top}r^{-}}{\partial p^{*}(f_{\theta})}\frac{\partial p^{*}(f_{\theta})}{\partial f_{\theta}}\frac{\max[p^{*}(f_{true})-p^{*}(f_{\theta}),0]}{p^{*}(f_{true})-p^{*}(f_{\theta})}\right. \\& \left. 
+ \phi^{-\textrm{ext}}\frac{\partial c^{\textrm{MCP,ext}\top}s^{-}}{\partial s^{*}(f_{\theta})}\frac{\partial s^{*}(f_{\theta})}{\partial f_{\theta}}\frac{\max[s^{*}(f_{true})-s^{*}(f_{\theta}),0]}{s^{*}(f_{true})-s^{*}(f_{\theta})}\right\}\frac{\partial f_{\theta}}{\partial \theta}.
\end{split}
\label{{eq:regret}G}
\end{equation}
The terms 
$\frac{\partial c^{\textrm{MCP}\top}r^{-}}{\partial p^{*}(f_{\theta})}$, 
$\frac{\partial c^{\textrm{MCP}\top}r^{+}}{\partial p^{*}(f_{\theta})}$, 
$\frac{\partial c^{\textrm{MCP,ext}\top}s^{-}}{\partial s^{*}(f_{\theta})}$, and 
$\frac{\partial c^{\textrm{MCP,ext}\top}s^{+}}{\partial s^{*}(f_{\theta})}$ in~(\ref{{eq:regret}G}) 
are straightforward to calculate, while the term $\frac{\partial f_{\theta}}{\partial \theta}$ is evaluated using the backpropagation algorithm. 
The terms $\frac{\partial p^{*}(f_{\theta})}{\partial f_{\theta}}$ and $\frac{\partial s^{*}(f_{\theta})}{\partial f_{\theta}}$ 
are obtained from an interior point (IP) based sub-gradient calculation, as proposed in~\cite{b27,b28} (the python code for interior point algorithm was developed using ChatGPT).
\subsubsection{Regret function design for DCOPF}\label{sec:regretdcopf}\noindent
The regret function for DCOPF is designed to train the load and the PTDF. The load training is based on the similar concept of generator ramping-up and -down. The load inaccuracy causes infeasibility in the solution, which therefore has a ramping-up/-down penalty for correction in real-time with ramping-up penalty greater than the ramping-down penalty.
\begin{figure*}
    \includegraphics[width=1.1\linewidth]{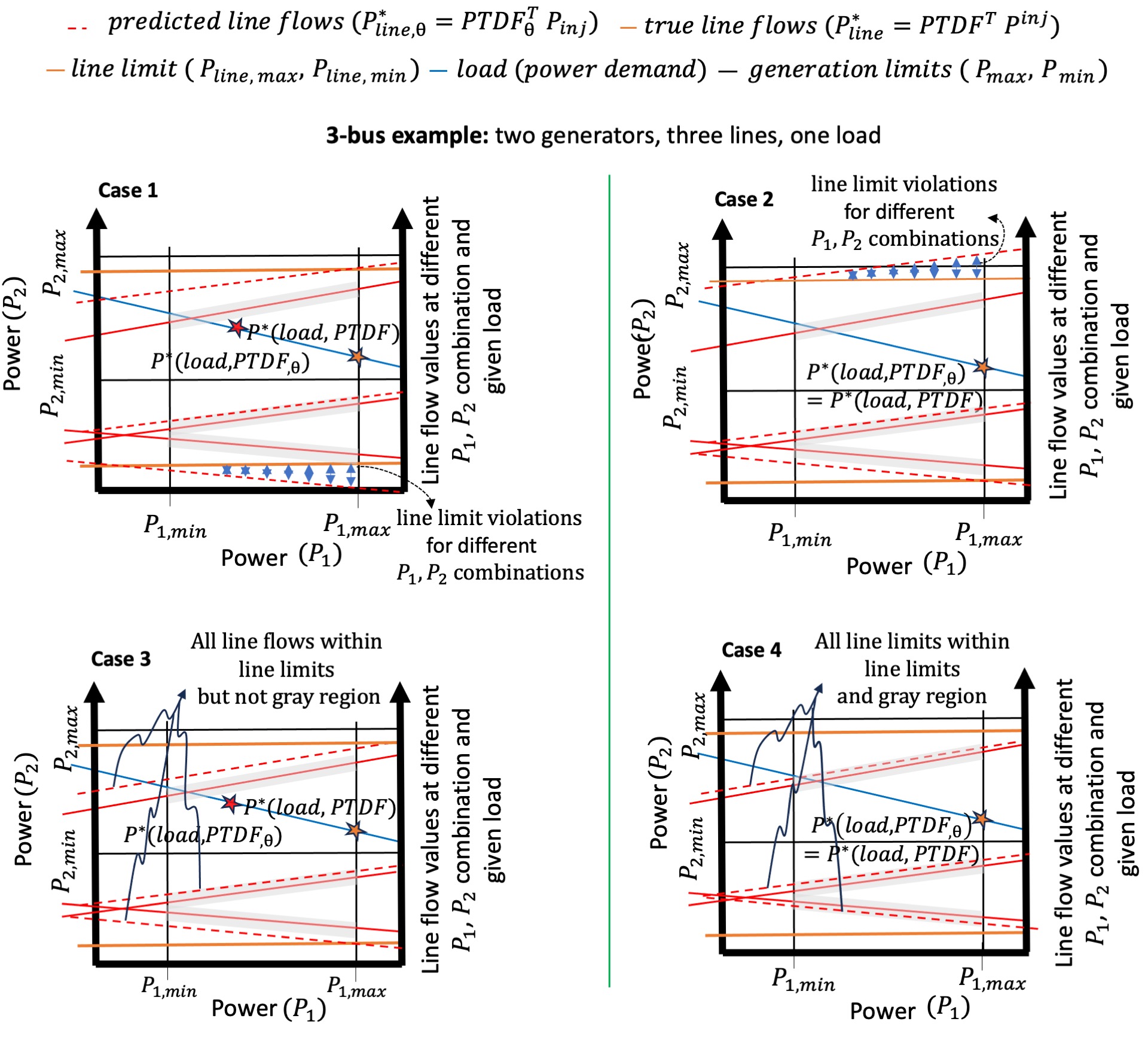}
    \caption{The four cases for feasible region with line flows representing impact of infeasibility in line flows on optimal solution for a given load.}
    \label{fig:cases_feas}
\end{figure*}

\begin{figure}
    \includegraphics[width=1\linewidth]{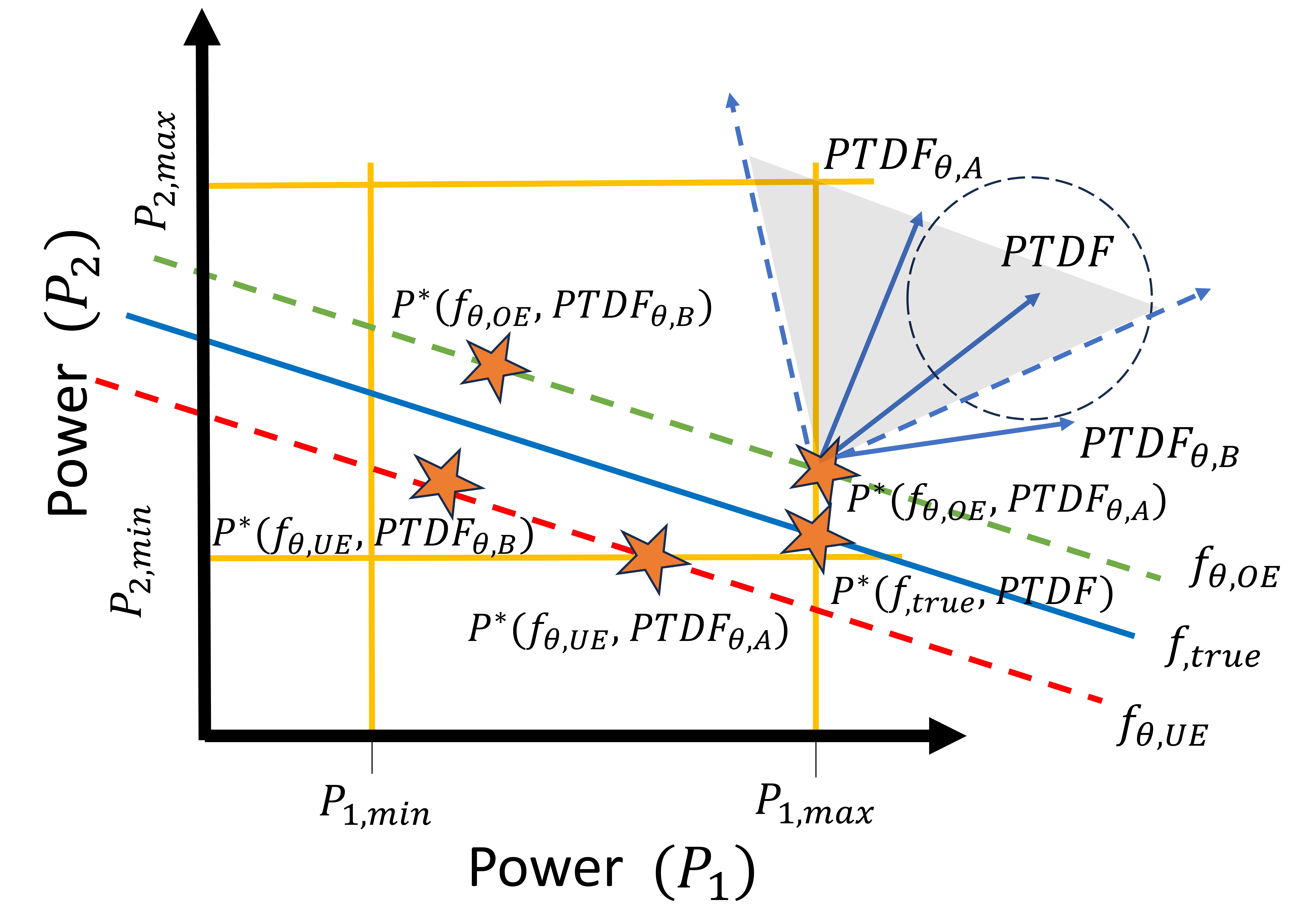}
    \caption{Feasible region for a two generator example. $f_{\theta,UE}$, $f_{true}$ and $f_{\theta,OE}$ represent the predicted underestimated, true and predicted overestimated load lines, respectively. The true solution $p^{*}(f_{true},PTDF)$ corresponds to true load and PTDF and also represent ED solution for $f_{true}$. This is just to show a true load reference and is not included in the sensitivity analysis example. The solution $p^{*}(f_{\theta,UE},PTDF)$ and $p^{*}(f_{\theta,OE},PTDF)$ represent optimum solutions corresponding to the under and overestimated loads ($f_{\theta,UE}$ and $f_{\theta,OE}$) respectively. The $f_{\theta,UE}$ and $f_{\theta,OE}$ cause the solution to go infeasible with respect to $f_{true}$ irrespective of the PTDF value. The $PTDF_{\theta,A}$ and $PTDF_{\theta,B}$ are two PTDF estimations. Solution corresponding to $PTDF_{\theta,A}$ ($p^{*}(f_{\theta,OE},PTDF_{\theta,A})$) represents the optimal solution (ED solution) corresponding to $f_{\theta,OE}$. The solution corresponding to $PTDF_{\theta,B}$ ($p^{*}(f_{\theta,OE},PTDF_{\theta,B})$) is the sub-optimum solution for $f_{\theta,OE}$ despite $PTDF_{\theta,A}$ and $PTDF_{\theta,B}$ being equidistant from the true PTDF. The gray region represents a sensitive region such that any solution corresponding to PTDF predictions within this region will be optimum. $p^{*}(f_{\theta,OE},PTDF_{\theta,A})$ was estimated using ILO while $p^{*}(f_{\theta,OE},PTDF_{\theta,B})$ was estimated using SLO. The similar analysis holds for $f_{\theta, UE}$ with respective solutions $p^{*}(f_{\theta,UE},PTDF_{\theta,A})$ and $p^{*}(f_{\theta,UE},PTDF_{\theta,B})$ corresponding to $PTDF_{\theta, A}$ and $PTDF_{\theta, B}$ respectively. For figure simplicity and clarity, the gray region is not shown for the underestimation case analysis.}
    \label{fig:case2_feas}
\end{figure}

The PTDF matrix inaccuracy as well due to being in equality constraints causes infeasibility in the solution. The infeasibility due to the inaccuracy of PTDF means that the prediction-driven line flows differ from the true line flows (line flows for which DCOPF generates ED solutions). In other words, $p^{line,\theta} = T_{\theta}(Mp + Ns -d)\neq p^{line,true}$, where $p^{line,\theta}$ represent prediction-driven line flows, $p^{line,true}$ represent line flows for which the DCOPF solution gives ED solution for a given load $d$. However, after thoroughly analyzing the feasible region for DCOPF, we found that the infeasibiliy of prediction-driven line flows with respect to true line flows will be mainly of four types illustrated in Fig.~\ref{fig:cases_feas}. 

In the first case, the infeasibility cause predicted line flows to violate line limits and true flows and DCOPF is sub-optimal with respect to (w.r.t.) ED. The sub-optimality w.r.t. ED implies a DCOPF generated solution less optimal (higher generator operation cost) compared to ED generated solution.
In the second case, predicted line flows violate line limits and differs from true line flows and DCOPF solution is optimal with respect to ED (DCOPF generate ED equivalent solution). In the third case, the predicted line flows are within line limits but differs from true line flows. Moreover, the line flows also exceed the gray region nearby true flows thereby giving DCOPF solutions less optimal than ED solutions. The gray region represents a range of line flows within which the solution of DCOPF equals the ideal/ED solution. In the fourth case, similar to the third case, only true line flows are violated. However, the predicted line flows are within the gray region thereby giving DCOPF solutions equivalent to ED solutions. 

In this work, we develop ILO formulations to obey the fourth case, where DCOPF give ED equivalent solutions while being within line limits. In other words, the ILO is formulated to learn DCOPF unknown parameters to produce line flows which are infeasible w.r.t. true line flows ($p^{line,\theta} \neq p^{line,true}$) constrained within line limits the corresponding dispatch solution to which is equivalent to ED solutions (represents ideal or no line congestion solution). PTDF is chosen as one of the learnable matrix in ILO formulation as it will achieve the aforementioned task due to its coefficients capturing system topology including line and bus configurations, and system size. Moreover, PTDF coefficients based on the system topology represent and can control power flow amounts at one line with respect to other lines. However, provided the structure of PTDF matrix coefficients capturing various aspects of system information, the PTDF matrix cannot be directly estimated using NN and require transformations on NN output to capture different abovementioned system information. The PTDF matrix coefficients are generated using PTDF mathematical formulations as a function of line impedances to represent the system topology. The NN output in the current work typically represents line impedances which are transformed to PTDF matrices using PTDF mathematical formulations.

The PTDF estimated to produce line flows within a certain PTDF magnitude range different than true line flows but within line limits to generate solutions approximating ED solutions will significantly minimize generator operational costs ahead of real-time and also real-time correction costs. This idea is further illustrated in Fig.~\ref{fig:case2_feas} using the feasible region of a simple two generator example. The example in Fig.~\ref{fig:case2_feas} explains the advantage of ILO over SLO for training PTDF. Assuming the current solution to be $p^{*}(f_{\theta,A},PTDF_{\theta,B})$ which is both infeasible w.r.t $f_{true}$ and sub-optimal/less optimal w.r.t $p^{*}(f_{\theta,A},PTDF)$ (also ED solution) due to inaccurate load and PTDF predictions. The correction for this solution will need correction for both infeasibility and sub-optimality. Assuming $p^{*}(f_{\theta,A},PTDF_{\theta,B})$ is corrected to optimality $p^{*}(f_{true},PTDF)$, the correction/regret function will be the difference between the cost of $p^{*}(f_{\theta,A},PTDF_{\theta,B})$ and $p^{*}(f_{true},PTDF)$. Moreover, since ramping-down is preferred more than ramping-up, the regret function will again be two terms with high penalty for ramping-up and low penalty for ramping-down. 
Overall, the regret function will correct for infeasibility due to inaccurate load to minimize RTM costs and sub-optimality due to inaccurate PTDF matrix to minimize RTM costs and line congestion to train load and PTDF matrix.
\begin{figure}[h!]
    \centering
    \includegraphics[width=0.9\linewidth]{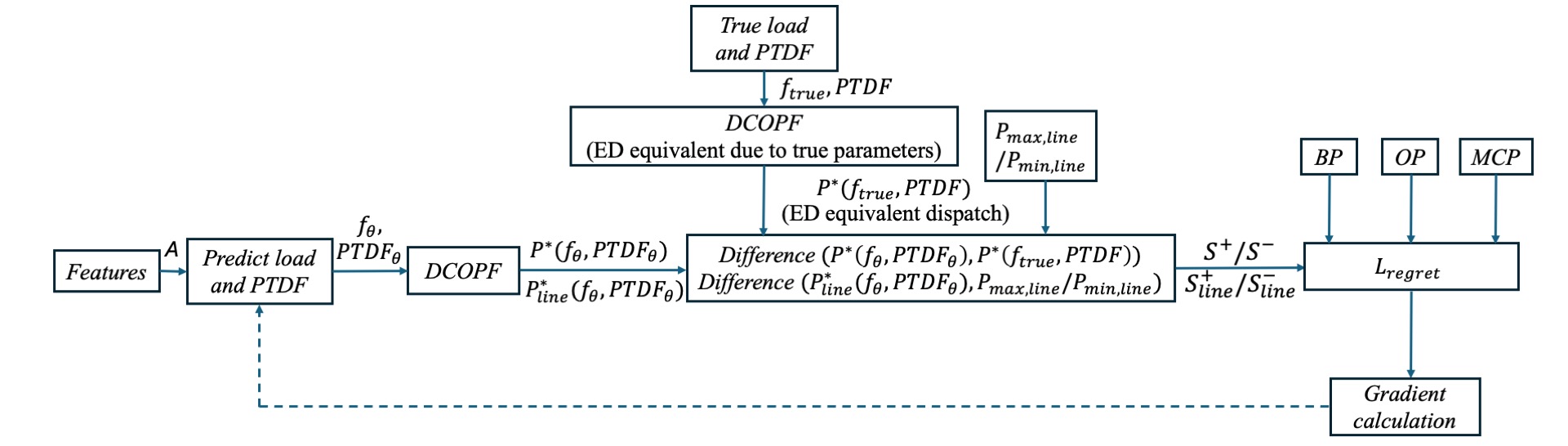}
    \caption{Load and PTDF training using ILO. The predicted load ($f_{\theta}$), predicted PTDF ($PTDF_{\theta}$) and true load ($f_{true}$), true PTDF ($PTDF$) are used to solve the DCOPF problem to generate corresponding power set-points $p^{*}(f_{\theta}, PTDF_{\theta})$ and $p^{*}(f_{true}, PTDF)$ respectively. The difference between the estimated and true power set-points along with the bidding price (BP), offer price (OP), and market clearing price (MCP)) are used to calculate the $L$ using~(\ref{{eq:regret}DCOPF}). The gradient to~(\ref{{eq:regret}DCOPF}) is calculated using using similar procedures as in Section 4 and backward passed to update $f_{\theta}$ and $PTDF_{\theta}$ parameters.}
    \label{fig:ILO_DCOPF}
\end{figure}
\begin{equation}
\begin{split}
L & := \phi^{+}c^{\textrm{MCP}\top}r^{+}
	+ \phi^{+\textrm{ext}}c^{\textrm{MCP,ext}\top}s^{+}
	+ \phi^{-}  c^{\textrm{MCP}\top}r^{-}
	+ \phi^{-\textrm{ext}}c^{\textrm{MCP,ext}\top}s^{-} \\
	&+ 1^{\top}f_{true}^{+} + 1^{\top}f_{true}^{-}\\
& \textrm{with the following constraints}\\
& r^{+}\geq p^{*}(f_{\theta},PTDF_{\theta})-p^{*}(f_{true},PTDF),\\
& r^{-}\geq p^{*}(f_{true},PTDF)-p^{*}(f_{\theta},PTDF_{\theta}),\\
& s^{+}\geq s^{*}(f_{\theta},PTDF_{\theta})-s^{*}(f_{true},PTDF),\\
& s^{-}\geq s^{*}(f_{true},PTDF)-s^{*}(f_{\theta},PTDF_{\theta}),\\
& d^{+}\geq f_{\theta, OE}-f_{true},\\
& d^{-}\geq f_{true} - f_{\theta, UE},\\
& r^{+}, \, r^{-}\geq 0,\\
& s^{+},\, s^{-}\geq 0, \\
& d^{+},\, d^{-} \geq 0,
\end{split}%
\label{{eq:regret}DCOPF}
\end{equation}%


The last two terms are designed as regularization to preserve load distribution at different nodes as a significant deviation in prediction compared to the true load at a specific line can be impractical. 
\subsubsection{Gradient for the DCOPF regret function}\label{sec:dcopfgradient}\noindent
For unknown parameter training in DCOPF, again the interior point based gradient calculation technique is used. The interior point (IP) objective function for this case contain $PTDF$ matrix with dimensions of $N_{L}\times N_{B}$ where $N_{B}$ denotes the number of buses. For the IEEE-14 system with 20 lines and 14 buses, the PTDF matrix size (20$\times$14) is very large for gradient calculation. This is due to the PTDF coefficients in the matrix as explained previously depend upon system topology including system size, line and bus placements, and interdependency of one line over all other lines and does not follow a regular pattern that can have regular gradients. Thus, the expressions in the gradient matrix for the regret function of this case study grows significantly large despite simplications and cannot be shown in the manuscript and can be only stored as a function inside the software. Nevertheless, the idea of IP based gradient calculations works for this case as well.
\section{Results}\label{sec:results}\noindent
Two case studies are considered along with the capability of ILO for congestion minimization. The two case studies are as follows:
\begin{itemize}
	\item Case study 1 - The ED of a power system including five generators with operating costs of 300, 400, 500, 600, and 700\$/MW, and maximum capacities of 2, 4, 3, 5, and 6 kW, respectively, is considered. The load demand is considered a contextual parameter that needs to be predicted for each hour.  The hour and 24-hour ahead load predictions were trained using the ED optimization model to minimize extra costs on demand participants at each hour and for the next 24 hours respectively. The load and its features data are taken from the independent system operator New England (ISONE) website corresponding to eight load zones. 
	\item Case study 2 - The DCOPF is applied to a IEEE-14 bus system with seven generators and eight loads. The seven generators in the system have the operating costs of 10, 20, 30, 40, 50, 60, and 70\$/MW. The maximum capacities of each generator are 60, 70, 40, 40, 50, and 20 MW respectively. The PTDF matrix and hour-ahead load predictions were trained to minimize line congestion and ramping costs. 
    
    Finally, the advantage of ILO training over SLO training in congestion minimization is shown and related with Fig.~\ref{fig:case2_feas}.
    
\end{itemize}



The PTDF matrix coefficients are estimated by governing the NN output (represent line impedance) using transformations to capture the system topology and line impedance interdependencies. The NN output provide values for line impedances which are governed to capture line impedance interdependencies and system topology using certain transformations and feedback training from the captured system processes in the overall ILO pipeline. IEEE-14 bus system with seven generators and eight loads is chosen in this case study. The historical data to train eight loads is obtained from eight different load zones of the independent system operator New England (ISONE) website. The features for PTDF learning are the same as the load  as the line congestion which is one of the objectives of this work depends on the value of load at different buses. The seven generators in the system have the operating costs of 10, 20, 30, 40, 50, 60, and 70\$/MW. The maximum capacities of each generator are 60, 70, 40, 40, 50, and 20 MW respectively. The python code for bar plots and tsne plots was developed using ChatGPT.
\subsection{Training setup}\noindent
A NN with three hidden layers each having 25 neurons is used for both the case studies. For both ED and DCOPF solution, the interior point optimization (IPOPT) solver~\cite{b29} was used. For model training, the pytorch and functorch libraries in python were used. In case study one, for the hour ahead case, the learning rates for ILO and SLO were set as $5\times10^{-3}$ and the barrier coefficient ($\mu$) was unbounded and was typically $1\times10^{-9}$ after solving through IPOPT. For the 24-hour ahead case, the learning rates for ILO and SLO were set as $1\times10^{-4}$ and $1\times10^{-3}$ and the barrier coefficient ($\mu$) was limited to $1\times10^{-7}$. In case study two, the learning rates for ILO were chosen as $3\times10^{-4}$ and $4\times10^{-3}$ respectively for load and PTDF training while for SLO load and PTDF training the learning rates were $5\times10^{-3}$ and $2\times10^{-3}$ respectively. 

In case study 1, the models are trained for different over and underestimation penalty parameters ($\phi^{-}_{i}$ and $\phi^{+}_{i}$ respectively, $ i\in$ total number of generators $(N_{G})$) to quantify different penalty impacts on model training and $L$. The penalty parameter settings are represented in Table~1.
\noindent
The first penalty setting corresponds to $\phi^{-}_{i}=\phi^{+}_{i}=1$ which practically represents the ramp-up and ramp-down prices (BP and OP respectively) are equal to MCP. In other words, the regulation market participants will not gain profit from providing regulation services if $\phi^{-}_{i}=\phi^{+}_{i}=1$. 
However, practically, producers/consumers willing to produce/consume extra to maintain real-time supply-demand balance would want to sell electricity at a price higher than MCP and buy electricity at less than MCP to maximize their profit.
Thus, for the remaining cases, the penalty terms denote the percentage of extra money concerning MCP paid by the demand participant under study for incorrect load estimations. The value of $\phi^{+}_{i}$ $>$ $\phi^{-}_{i}$ denotes higher ramp-up costs compared to ramp-down costs. Moreover, except for the first case ($\phi^{+}_{i}$ = $\phi^{-}_{i}$ = 1), the $\phi^{+}_{i}$ and $\phi^{-}_{i}$ settings in Table~1 represent growing differences between $\phi^{+}_{i}$ and $\phi^{-}_{i}$ to show the impact of higher differences between ramp-up and ramp-down costs on ILO training. 

In case study 2, the cost vector in the regret function is chosen to be the same as the generator operating costs. This is due to the objective of congestion minimization which requires expensive generators to receive higher penalties. The ramping-up and -down penalties respectively for deviations from schedule are setting1: $\phi^{-}_{1-N}$ = 1, $\phi^{+}_{1-N}$ = 1, setting2: $\phi^{-}_{1-N}$ = 1.02, $\phi^{+}_{1-N}$ = 1.06, setting3: $\phi^{-}_{1-N}$ = 1.05, $\phi^{+}_{1-N}$ = 1.1, setting4: $\phi^{-}_{1-N}$ = 1.08, $\phi^{+}_{1-N}$ = 1.15, and setting5: $\phi^{-}_{1-N}$ = 1.12, $\phi^{+}_{1-N}$ = 1.22, where $N$ is the number of generators. Since, the generator operating costs are used as cost vectors assuming the pricing order of geneators in RTM will remain the same as during hour-ahead scheduling, each parameter setting is equally distributed over all the generators.

The ILO training results for both case studies are compared with the SLO in terms of $L$ value in order to compare their performance. Moreover, for case study 2, the performance of ILO for approximating sub-optimal DCOPF problem to generate ED solutions (DCOPF solution using true PTDF matrix) is compared with SLO by comparing their operational costs. The improved regret function and operational costs indicates significantly enhanced real-time market operations and hour-ahead generator scheduling. 

\begin{table*}[!h]
\centering
\begin{scriptsize}
\label{tab1}
\caption{Penalty parameter settings for all grid generators (G1-G5)}
\begin{tabular}{|c|cc|cc|cl|ll|ll|}
\hline
\multirow{2}{*}{\textbf{\begin{tabular}[c]{@{}c@{}} Penalty parameter\\ setting\end{tabular}}} & \multicolumn{2}{c|}{\textbf{G1}}                               & \multicolumn{2}{c|}{\textbf{G2}}                               & \multicolumn{2}{c|}{\textbf{G3}}                                                    & \multicolumn{2}{c|}{\textbf{G4}}                                                    & \multicolumn{2}{c|}{\textbf{G5}}                                                    \\ \cline{2-11} 
                                                                                      & 
\multicolumn{1}{c|}{$\phi^{-}$} & $\phi^{+}$ & \multicolumn{1}{c|}{$\phi^{-}$} & $\phi^{+}$ &  \multicolumn{1}{c|}{$\phi^{-}$} & $\phi^{+}$ &  \multicolumn{1}{c|}{$\phi^{-}$} & $\phi^{+}$ &  \multicolumn{1}{c|}{$\phi^{-}$} & $\phi^{+}$   \\                                                                                      
%
                                                                                      \hline
Setting 1                                                                             & \multicolumn{1}{c|}{1}               & 1                  & \multicolumn{1}{c|}{1}               & 1              & \multicolumn{1}{c|}{1}                 &    1                                      & \multicolumn{1}{l|}{1}                    &       1                                   & \multicolumn{1}{l|}{1}                    &             1                             \\ \hline
Setting 2                                                                             & \multicolumn{1}{c|}{1.05}               & 1.1                  & \multicolumn{1}{c|}{1.1}               & 1.2               & \multicolumn{1}{c|}{1.15}                 &  1.25                                        & \multicolumn{1}{l|}{1.2}                    &  1.3                                        & \multicolumn{1}{l|}{1.25}                    &  1.35                                        \\ \hline
Setting 3                                                                             & \multicolumn{1}{c|}{1.1}               & 1.2                  & \multicolumn{1}{c|}{1.15}               & 1.3               & \multicolumn{1}{c|}{1.2}                 &  1.35                                        & \multicolumn{1}{l|}{1.25}                    &       1.4                                   & \multicolumn{1}{l|}{1.3}                    &    1.45                                      \\ \hline
Setting 4                                                                             & \multicolumn{1}{c|}{1.2}                & 1.35                  & \multicolumn{1}{c|}{1.25}                 & 1.45                & \multicolumn{1}{c|}{1.3}                 &     1.55                                     & \multicolumn{1}{l|}{1.35}                    &   1.65                                       & \multicolumn{1}{l|}{1.4}                    &   1.75                                       \\ \hline
Setting 5                                                                             & \multicolumn{1}{c|}{1.35}               & 1.65                & \multicolumn{1}{c|}{1.4}               & 1.75                & \multicolumn{1}{c|}{1.45}                 &   1.85                                       & \multicolumn{1}{l|}{1.5}                    &  1.95                                        & \multicolumn{1}{l|}{1.55}                    &     2.05                                     \\ \hline
\end{tabular}
\end{scriptsize}
\end{table*}

\subsection{Case Study 1}\noindent 
In the hour-ahead case, the NN output is a single neuron representing the hour-ahead load prediction corresponding to the environmental features/context. The predictions trained with ILO are compared with the predictions trained with SLO to show the effectiveness of ILO in minimizing $L$ or minimizing ramping costs. Both models are trained using the load data for five days while the predictions are tested using the contextual information of the next two days. 
The training and testing results for case study 1 corresponding to all parameter settings (setting 1 - setting 5) are shown in Table~\ref{tab:rescase1}. 
\begin{table*}[!h]
\centering
\begin{scriptsize}
\label{tab2}
\caption{Training and testing results for ILO and SLO for hour-ahead load model}\label{tab:rescase1}
\begin{tabular}{|c|cc|c|cc|c|}
\hline
\multirow{2}{*}{\begin{tabular}[c]{@{}c@{}}Penalty parameter \\ setting\end{tabular}} & \multicolumn{2}{c|}{$\textbf{L}_{\textbf{regret}}$ $\textbf{ILO}$} & \multirow{2}{*}{\textbf{Epochs ILO}} & \multicolumn{2}{c|}{$\textbf{L}_{\textbf{regret}}$ $\textbf{SLO}$} & \multirow{2}{*}{\textbf{Epochs SLO}} \\ \cline{2-3} \cline{5-6}
                                                                                      & \multicolumn{1}{c|}{Training}  & Testing  &                                      & \multicolumn{1}{c|}{Training}  & Testing  &                                      \\ \hline
Setting 1                                                                             & \multicolumn{1}{c|}{\textbf{0.218}}     & \textbf{0.378}    & 100                                   & \multicolumn{1}{c|}{0.428}     & 0.646    & 250                                  \\ \hline
Setting 2                                                                             & \multicolumn{1}{c|}{\textbf{0.224}}     & \textbf{0.288}    & 100                                   & \multicolumn{1}{c|}{0.577}     & 0.863    & 250                                  \\ \hline
Setting 3                                                                             & \multicolumn{1}{c|}{\textbf{0.190}}      & \textbf{0.205}    & 100                                   & \multicolumn{1}{c|}{0.591}     & 0.854    & 250                                  \\ \hline
Setting 4                                                                             & \multicolumn{1}{c|}{\textbf{0.180}}      & \textbf{0.230}     & 100                                   & \multicolumn{1}{c|}{0.700}       & 1.040     & 250                                  \\ \hline
Setting 5                                                                             & \multicolumn{1}{c|}{\textbf{0.225}}     & \textbf{0.298}    & 100                                  & \multicolumn{1}{c|}{0.814}     & 1.130     & 250                                  \\ \hline
\end{tabular}
\end{scriptsize}
\end{table*}
As observed in Table~\ref{tab:rescase1}
, the ILO trained $L$ exhibit smaller values than SLO for both the training and testing instances for all the settings of $\phi^{-}_{i}$ and $\phi^{+}_{i}$. Moreover, with growing differences between ramping-up and -down costs, the $L$ for SLO increases faster than ILO. This is due to ILO model training focusing on minimizing extra costs on the demand participant, unlike SLO which focuses on minimizing load prediction error. 
%

In the 24-hour ahead load training, the load model is trained to estimate the next 24-hour load at a one-hour time grid. 
The training procedure for the 24-hour case is the same as for the one-hour case. The model is trained for 6 hours and tested for the next 3 hours. The training and testing $L$ for SLO and ILO for the 24-hour-ahead load model are shown in Table~\ref{tab:rescase2}.  
 The penalty parameters for the 24-hour case are set to be the same as for the one-hour case. It is observed from the results, for all the settings of penalty parameters the ILO based approach exhibit lower regret values compared to the SLO approach. The lower regret function indicates lower extra costs on the demand participants for deviation from scheduled demand.

\begin{table*}[!h]
\caption{Training and testing results for ILO and SLO for 24-ahead load model}\label{tab:rescase2}
\centering
\begin{scriptsize}
\begin{tabular}{|c|cc|c|cc|c|}
\hline
\multirow{2}{*}{\begin{tabular}[c]{@{}c@{}}Penalty parameter \\ setting\end{tabular}} & \multicolumn{2}{c|}{$\textbf{L}_{\textbf{regret}}$ $\textbf{ILO}$} & \multirow{2}{*}{\textbf{Epochs ILO}} & \multicolumn{2}{c|}{$\textbf{L}_{\textbf{regret}}$ $\textbf{SLO}$} & \multirow{2}{*}{\textbf{Epochs SLO}} \\ \cline{2-3} \cline{5-6}
                                                                                      & \multicolumn{1}{c|}{Training}  & Testing  &                                      & \multicolumn{1}{c|}{Training}  & Testing  &                                      \\ \hline
Setting 1                                                                             & \multicolumn{1}{c|}{\textbf{6.6}}      & \textbf{19.1}     & 100                                   & \multicolumn{1}{c|}{8.0}      & 19.3     & 100                                  \\ \hline
Setting 2                                                                             & \multicolumn{1}{c|}{\textbf{16.7}}      & \textbf{29.8}     & 100                                  & \multicolumn{1}{c|}{26.9}      & 35.1     & 100                                  \\ \hline
Setting 3                                                                             & \multicolumn{1}{c|}{\textbf{12.4}}     & \textbf{26.3}     & 100                                   & \multicolumn{1}{c|}{21.9}      & 31.2     & 100                                  \\ \hline
Setting 4                                                                             & \multicolumn{1}{c|}{\textbf{11.0}}     & \textbf{31.7}    & 100                                  & \multicolumn{1}{c|}{25.6}      & 37.6     & 100                                  \\ \hline
Setting 5                                                                             & \multicolumn{1}{c|}{\textbf{26.9}}     & \textbf{44.8}     & 100                                   & \multicolumn{1}{c|}{36.4}      & 53.5     & 100                                  \\ \hline
\end{tabular}
\end{scriptsize}
\end{table*}
\subsection{Case Study 2}\noindent 
In this case, the load and PTDF are trained at one hour time resolution. Provided the current feature data, the load and PTDF are predicted every next hour. 
The ILO training and testing results are then compared to SLO training and testing results with the objective of minimizing the regret function. 
Table~\ref{tab:rescase2b} illustrates the comparison between regret functions corresponding to ILO and SLO training. As observed, for all the settings of penalty parameters the regret function of ILO remains smaller than that of SLO. The smaller regret function, as explained previously, enhances economic operation by minimizing real-time correction costs and hour-ahead operational costs of generators.
\begin{table*}[!h]
\centering
\begin{scriptsize}
\caption{Training and testing results for ILO and SLO for hour-ahead load and PTDF model}\label{tab:rescase2b}
\begin{tabular}{|c|cc|c|cc|c|}
\hline
\multirow{2}{*}{\begin{tabular}[c]{@{}c@{}}Penalty parameter \\ setting\end{tabular}} & \multicolumn{2}{c|}{$\textbf{L}_{\textbf{regret}}$ $\textbf{ILO}$} & \multirow{2}{*}{\textbf{Epochs ILO}} & \multicolumn{2}{c|}{$\textbf{L}_{\textbf{regret}}$ $\textbf{SLO}$} & \multirow{2}{*}{\textbf{Epochs SLO}} \\ \cline{2-3} \cline{5-6}
                                                                                      & \multicolumn{1}{c|}{Training}  & Testing  &                                      & \multicolumn{1}{c|}{Training}  & Testing  &                                      \\ \hline
Setting 1                                                                             & \multicolumn{1}{c|}{\textbf{2015}}     & \textbf{2463}    & 100                                   & \multicolumn{1}{c|}{5188}     & 4557    & 100                                  \\ \hline
Setting 2                                                                             & \multicolumn{1}{c|}{\textbf{1751}}     & \textbf{2926}    & 100                                   & \multicolumn{1}{c|}{4352}     & 3835    & 100                                  \\ \hline
Setting 3                                                                             & \multicolumn{1}{c|}{\textbf{2375}}      & \textbf{2997}    & 100                                   & \multicolumn{1}{c|}{4775}     & 4293   & 100                                  \\ \hline
Setting 4                                                                             & \multicolumn{1}{c|}{\textbf{2380}}      & \textbf{2801}     & 100                                   & \multicolumn{1}{c|}{4336}       & 4202     & 100                                  \\ \hline     
Setting 5                                                                             & \multicolumn{1}{c|}{\textbf{1310}}      & \textbf{1656}     & 100                                   & \multicolumn{1}{c|}{4846}       & 4596     & 100                                  \\ \hline     
\end{tabular}
\end{scriptsize}
\end{table*}

The ILO trains load to be either more overestimate than underestimate or train underestimate load with high accuracy while SLO trains for accuracy which may be over/underestimate and is different from ILO. For PTDF, the ILO trains the PTDFs to be within the gray region explained in Fig.~\ref{fig:case2_feas} to obtain better optimal solutions. 

\subsection{ILO for Congestion Minimization}
\begin{figure}
    \centering
    \includegraphics[width=1\linewidth]{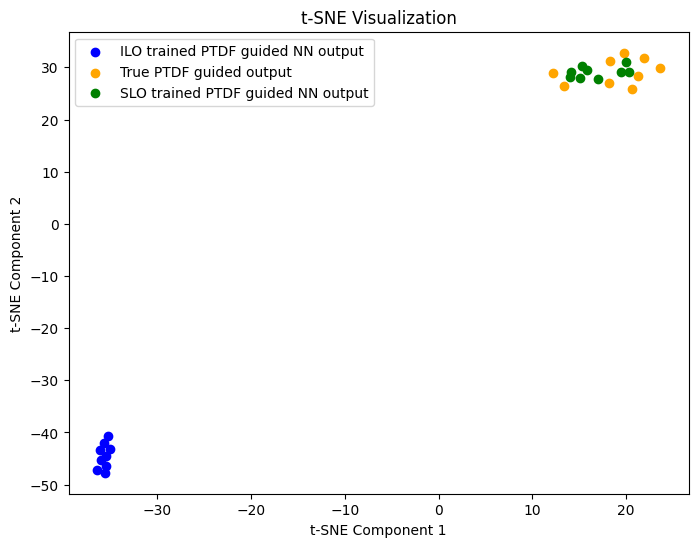}
    \caption{t-SNE analysis for PTDFs learned using ILO, and SLO with true PTDF as a reference. The plot maps the 20-D PTDF guided NN vector for each data point on the 2-D plot.}
    \label{fig:tsne}
\end{figure}

\begin{figure}
    \centering
    \includegraphics[width=1\linewidth]{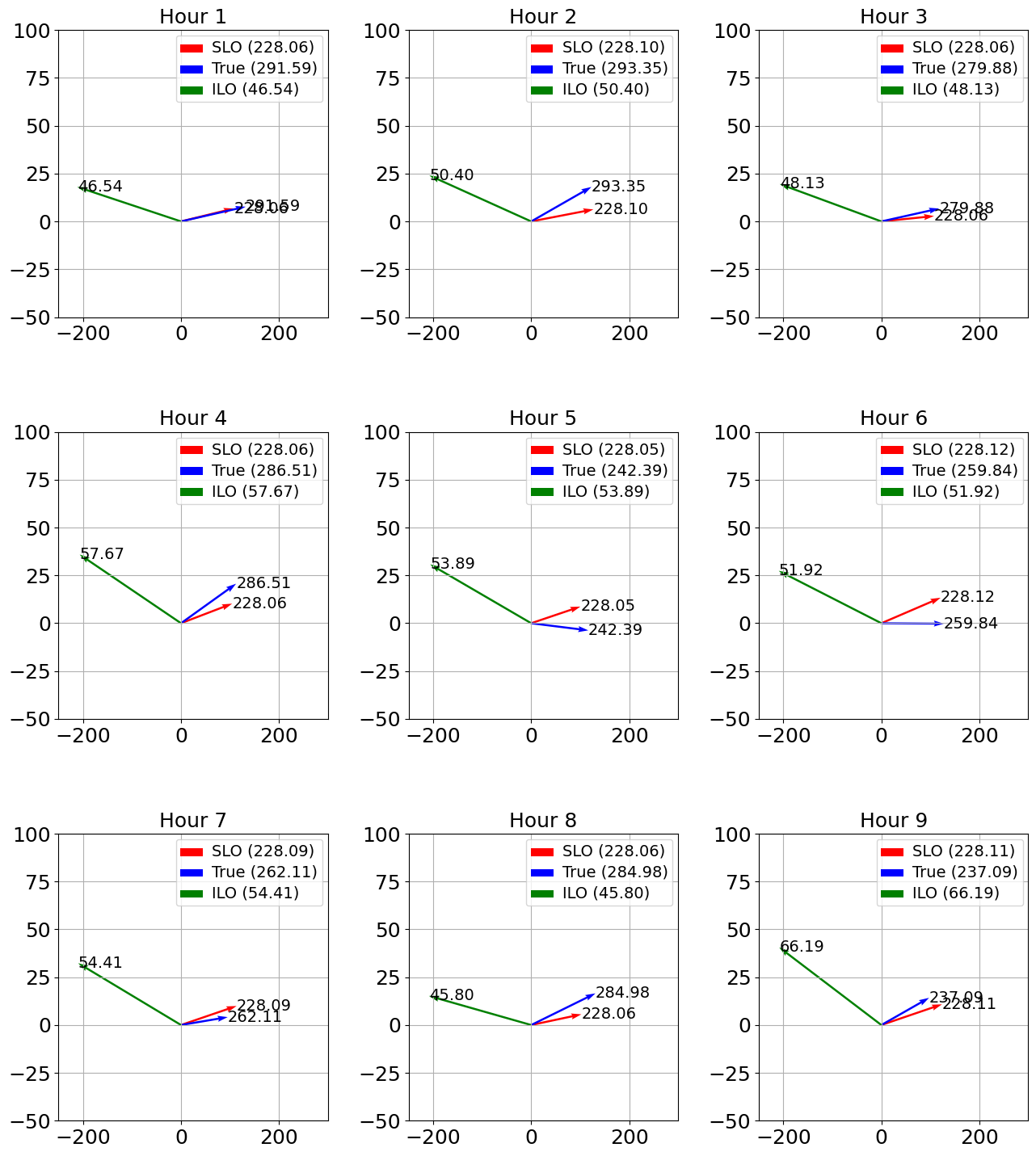}
    \caption{Magnitude and direction plot over 9 hour operation for PTDF guided SLO, true and ILO trained NN vectors. The magnitude is the direct magnitude of the 20-D vector for each data point, while the angle is the tsne based dimensionality reduction.d}
    \label{fig:tsne_magang}
\end{figure}

To show the ILO capability in approximating sub-optimal DCOPF solutions to optimal ED solutions (congestion minimization) compared with SLO, the PTDF training for both ILO and SLO is compared for same load. 
To understand the advantage of ILO based training for congestion minimization, recall the two generator feasible region example explained in Fig.~\ref{fig:case2_feas}. The estimated PTDFs, namely $PTDF_{\theta, A}$ and $PTDF_{\theta, B}$ were equidistant from the true $PTDF$. However, $PTDF_{\theta, A}$ being ILO trained and within the sensitivity range (gray region) generate ED solution while $PTDF_{\theta, B}$ being SLO trained and outside of the sensitivity range provided sub-optimal solution. Provided that, the training results for PTDFs of ILO and SLO are illustrated in Figs.~\ref{fig:tsne} and~\ref{fig:tsne_magang} to estimate their closeness to the true PTDF value which corresponds to ED solution. The plots mainly illustrates the direct 20-D NN output as 2-D using t-SNE for visualization which was guided to find correlations between line impedances given a system topology using transformations instead of plotting the transformed output. In Fig.~\ref{fig:tsne}, the 20-D values for each data point were directly plotted on a 2-D plot, while for Fig.~\ref{fig:tsne_magang} the magnitude represents the magnitude of the 20-D vector and the direction is the 2-D angular representation of 20-D vectors. It is clearly observed that the ILO based PTDFs are far from the true value and ILO does not train for accuracy. While SLO due to accurate training is much closer to the true PTDF value.
\begin{figure*}
    \centering
    \includegraphics[width=1\linewidth]{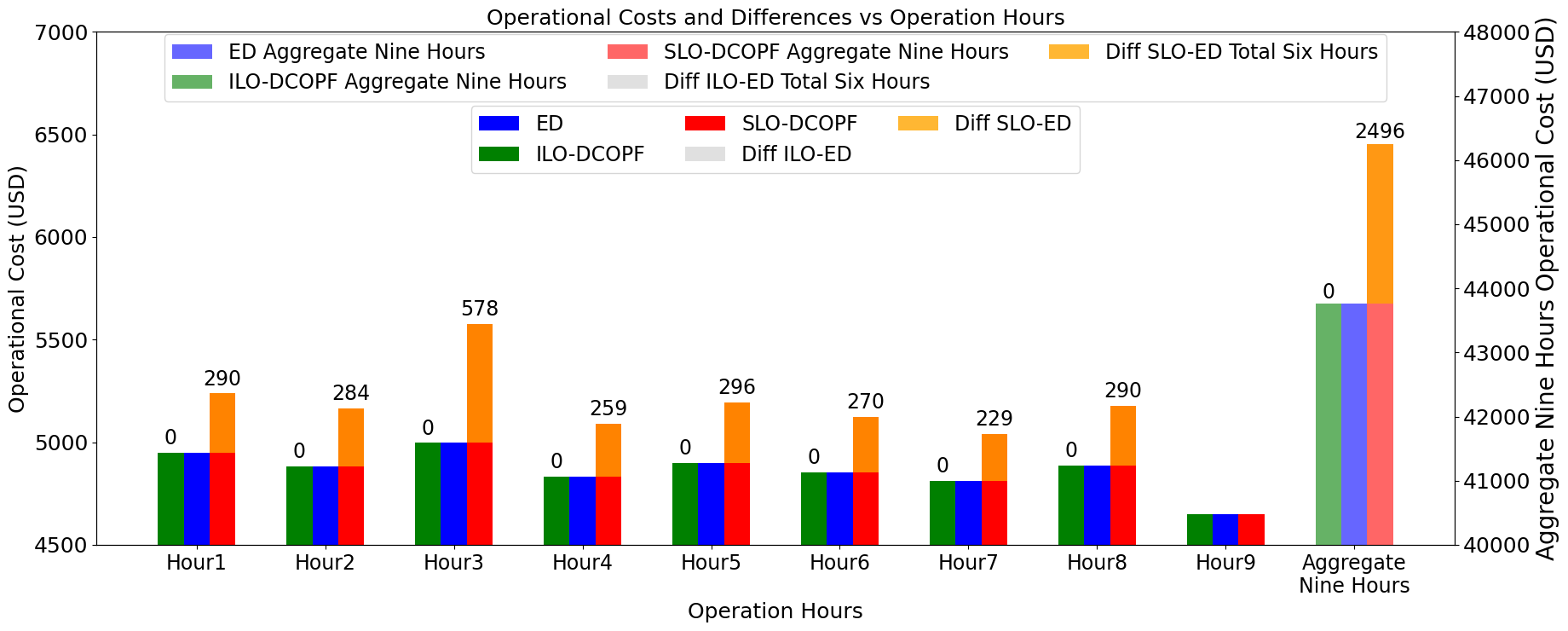}
    \caption{Hourly and total operational cost comparison between ILO and SLO for training.}
    \label{fig:ILO_Oper_Train}
\end{figure*}
\begin{figure*}
    \centering
    \includegraphics[width=1\linewidth]{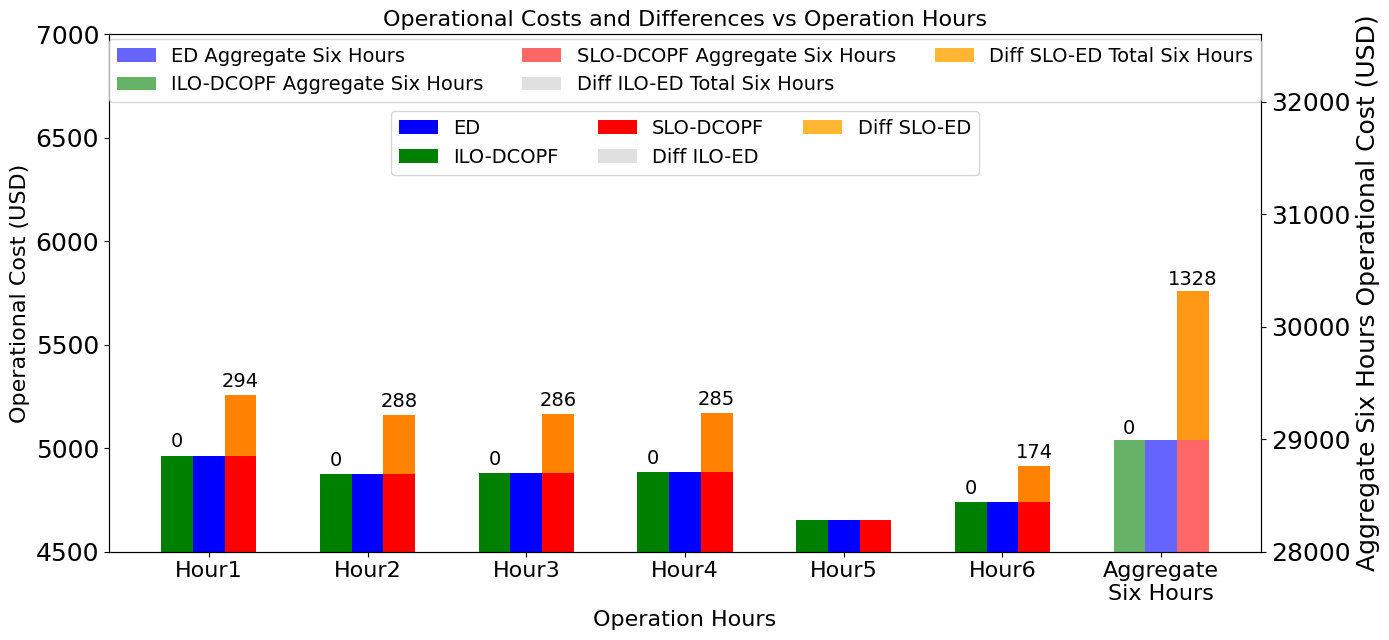}
    \caption{Hourly and total operational cost comparison between ILO and SLO for testing.}
    \label{fig:ILO_Oper_Test}
\end{figure*}
Nevertheless, the ILO trained PTDF generate ED solutions while SLO trained PTDF despite being significantly closer to the true values generate sub-optimal solutions as explained in the following figures. 

Figs.~\ref{fig:ILO_Oper_Train} and~\ref{fig:ILO_Oper_Test} illustrate the hourly and total operational costs of all the generators for training and testing instances respectively obtained using ILO and SLO. The training instances are shown for 9 hours while the testing instances for 6 hours. 

The results indicate ILO operational costs being equal to ED operational costs at all hours of operation for both training and testing instances. The trained PTDFs of ILO were much different compared to true PTDFs as shown in Fig.~\ref{fig:tsne} based on t-SNE (t-distributed Stochastic Neighbor Embedding) analysis while the trained PTDFs using SLO were very close to the true value; however, the ILO based DCOPF is still approximated to ED due to being within the sensitivity range as explained in Fig.~\ref{fig:case2_feas} while SLO despite being significantly closer to true PTDFs cannot obtain true ED solutions. 
\section{Conclusion}\label{sec:conclusions}\noindent
As a conclusion, the proposed ILO methodologies were compared with SLO for ED and DCOPF parameter training. For both case study 1 and 2 for all the settings the ILO outperformed SLO in terms of achieving a lower regret function. The load trained using ILO for both the case studies was more an overestimate than an underestimate to achieve a lower regret function. The real-time correction costs for incorrect load and PTDF estimations were thus minimized due to better regret function training as evident in the results. Moreover, for the second case study, the lower regret function is also an indication of lower operational costs in the hour ahead scheduling as the regret function corrects for both optimality and feasibility. 
\section*{Notation}
\noindent
\textbf{Sets}\newline
$\cb$ - Buses.\\
$\ci$ - Generation units. \\
$\cj$ - Loads.\\
$\cl$ - Transmission lines.\\
\textbf{Known parameters\footnote{Do not depend on the context.}}\\
\textbf{Unknown parameters\footnote{Depend on the context and need to be predicted as a function of the context.}}\\
\textbf{Continuous variables\footnote{Obtained as a solution to an optimization problem.}}\\
\section*{Abbreviations}\noindent
ED - Economic dispatch\\
ERT - Economic hybrid with reference tracking\\
DAM - Day-ahead market\\
DCOPF - DC optimal power flow\\
DRO - Decision rule optimization \\
DRL - Deep reinforcement learning \\
IDM - Intra-day market\\
ILO - Integrated learning and optimization\\
SPO+ - Smart predict then optimize \\
ISO - Independent system operator\\
LO - Learning and optimization\\
MO - Market operator\\
MPC - Model predictive control\\
MCP - Market clearing price \\
BP - Bidding price \\
OP - Offer price \\
NN - Neural network\\
CNN - Convolutional neural network \\
LSTM - Long short term memory \\
LM-BP -  Levenberg–Marquardt back-propagation \\
DCNN - Deep convolutional neural network \\
GP - Gaussian process \\
RF - Random forest \\
GB - Gradient boosting \\
DP - Dynamic programming \\
MC - Monte carlo \\
RWM - Roulette wheel mechanism \\
DR - Demand response \\
PDF - Probability distribution function \\
MCMC - Markov chain Monte Carlo \\
BRP - Balance responsible party \\
TFT - temporal fusion transformer \\  
PTDF - Power transfer distribution factor\\
RTM - Real-time market \\
SLO - sequential learning and optimization\\
ISONE - Independent system operator New England \\
IPOPT - Interior point optimization \\
IP - Interior point \\
LP - Linear program \\
RoCoF - Rate of change of frequency \\
EV - Electric vehicle \\

\section*{AI declaration statement}
\noindent
During the preparation of this work the author(s) used ChatGPT in order to develop the code for interior point algorithm for DCOPF and verified its performance by comparing it with the actual results. Moreover, the code for generating bar plots and tsne plots was also developed using chatGPT. After using this tool/service, the author(s) reviewed and edited the content as needed and take(s) full responsibility for the content of the publication.

\end{document}